%% file: main.tex
\documentclass{ieeeaccess}
\usepackage{amsmath,amsfonts}
\usepackage{algorithm}
\usepackage{algorithmic}
\usepackage{array}
\usepackage{subcaption}
\usepackage{textcomp}
\usepackage{stfloats}
\usepackage{url}
\usepackage{verbatim}
\usepackage{graphicx}
\usepackage{cite}
\usepackage{xcolor}
\newcommand{\sol}{\text{co-TenQu}}

\def\BibTeX{{\rm B\kern-.05em{\sc i\kern-.025em b}\kern-.08em
    T\kern-.1667em\lower.7ex\hbox{E}\kern-.125emX}}

\begin{document}

\title{A Quantum-Classical Collaborative Training Architecture Based on Quantum State Fidelity }


\author{\uppercase{
Ryan L'Abbate\authorrefmark{1}, 
Anthony D'Onofrio Jr.\authorrefmark{1}, 
Samuel Stein\authorrefmark{2},
Samuel Yen-Chi Chen\authorrefmark{3},
Ang Li\authorrefmark{2}, Pin-Yu Chen\authorrefmark{4}, Juntao Chen\authorrefmark{1},
Ying Mao\authorrefmark{1}}}
\address[1]{Computer and Information Science Department, Fordham University, New York, USA. E-mail: \{rlabbate, adonofrio10, jchen504, ymao41\}@fordham.edu}
\address[2]{Pacific Northwest National Laboratory, Richland, WA, USA. Email: \{samuel.stein, ang.li\}@pnnl.gov}
\address[3]{Brookhaven National Laboratory, Upton, NY, USA. Email:\{ycchen1989@gmail.com\}}
\address[4]{IBM Research, Yorktown Heights, NY, USA. Email:\{pin-yu.chen@ibm.com\}}

\markboth
{L'Abbate \headeretal: A Quantum-Classical Collaborative Training Architecture Based on Quantum State Fidelity}
{L'Abbate \headeretal: A Quantum-Classical Collaborative Training Architecture Based on Quantum State Fidelity}

\corresp{Corresponding author: Ying Mao (e-mail: ymao41@fordham.edu).}

\tfootnote{This research was supported in part by the National Science Foundation (NSF) under grant agreements 2329020, 2335788, and 2301884. It was also supported in part by the U.S. Department of Energy (DOE) through the Office of Advanced Scientific Computing Research's “Advanced Memory to Support Artificial Intelligence for Science”. PNNL is operated by Battelle for the DOE under Contract DE-AC05-76RL01830.}

\begin{abstract}
Recent advancements have highlighted the limitations of current quantum systems, particularly the restricted number of qubits available on near-term quantum devices. This constraint greatly inhibits the range of applications that can leverage quantum computers. Moreover, as the available qubits increase, the computational complexity grows exponentially, posing additional challenges. Consequently, there is an urgent need to use qubits efficiently and mitigate both present limitations and future complexities. To address this, existing quantum applications attempt to integrate classical and quantum systems in a hybrid framework.
In this study, we concentrate on quantum deep learning and introduce a collaborative classical-quantum architecture called co-TenQu. The classical component employs a tensor network for compression and feature extraction, enabling higher-dimensional data to be encoded onto logical quantum circuits with limited qubits. On the quantum side, we propose a quantum-state-fidelity-based evaluation function to iteratively train the network through a feedback loop between the two sides. co-TenQu has been implemented and evaluated with both simulators and the IBM-Q platform. Compared to state-of-the-art approaches, co-TenQu enhances a classical deep neural network by up to 41.72\% in a fair setting. Additionally, it outperforms other quantum-based methods by up to 1.9 times and achieves similar accuracy while utilizing 70.59\% fewer qubits.
\end{abstract}


\doi{10.1109/TQE.2024.3367234}

\maketitle

\input{sources/intro}

\input{sources/related}
\input{sources/motivation}

\input{sources/system}

\input{sources/evaluation}


\bibliographystyle{IEEEtran}
\bibliography{ref.bib}

\vskip -2.5\baselineskip plus -1fil
\begin{IEEEbiography}
[{\includegraphics[width=1in,height=1.25in,clip,keepaspectratio]{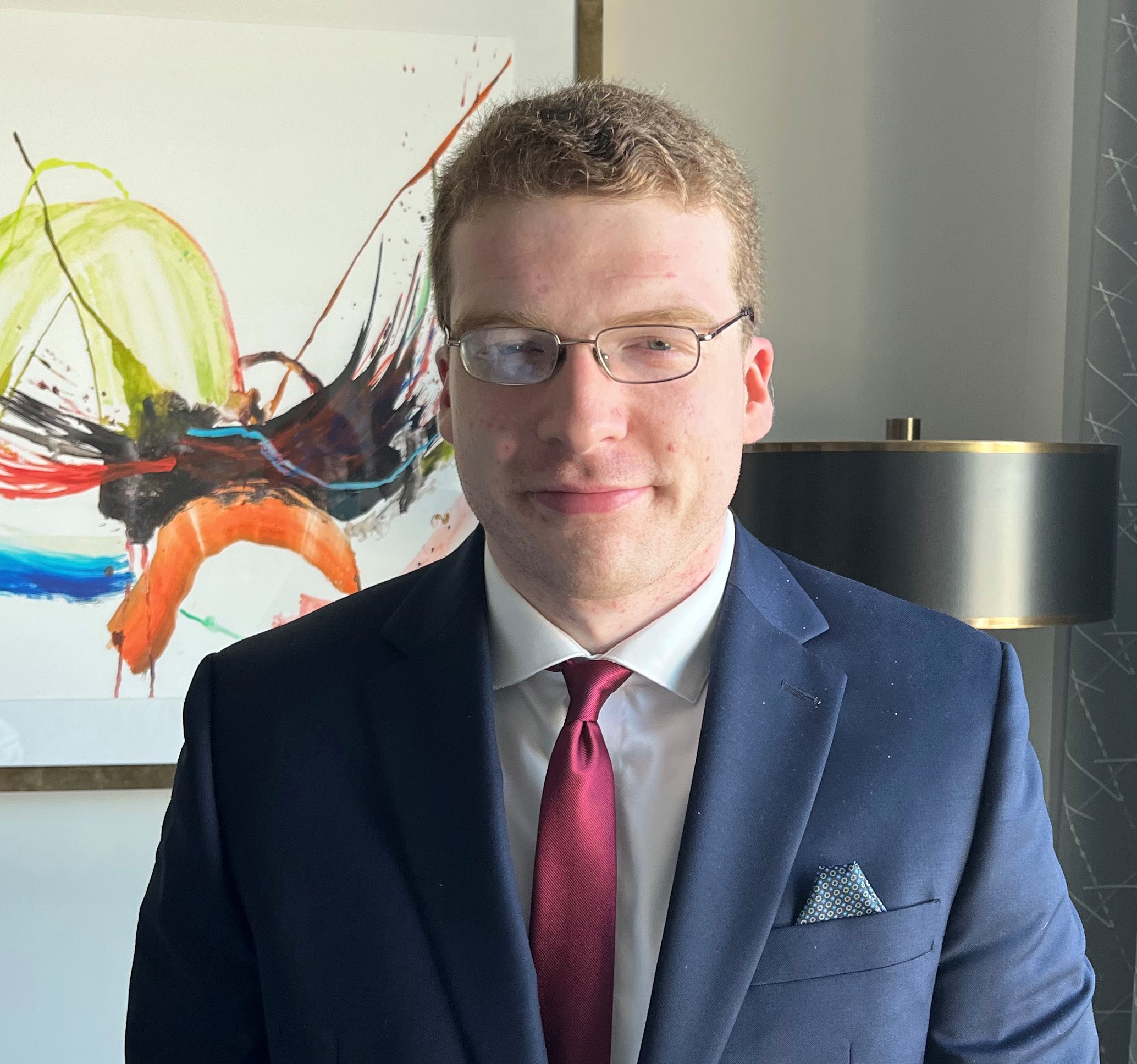}}]%
{Ryan L'Abbate} is a database developer and research chemist at En-Tech Corp. He received bachelor's degrees in Chemical Engineering and Mathematics from Manhattan College in December 2017 and a master's degree in Data Science from the Department of Computer and Information Science at Fordham University in May 2022. He was inducted into the Omega Chi Epsilon honors society for chemical engineering and the Pi Mu Epsilon honors society for mathematics. He has also had data science research published in the \emph{Haseltonia} journal. His research interests include quantum computing, quantum data science, and data structures.
\end{IEEEbiography}

\vskip -2.5\baselineskip plus -1fil

\begin{IEEEbiography}[{\includegraphics[width=1in,height=1.25in,clip,keepaspectratio]{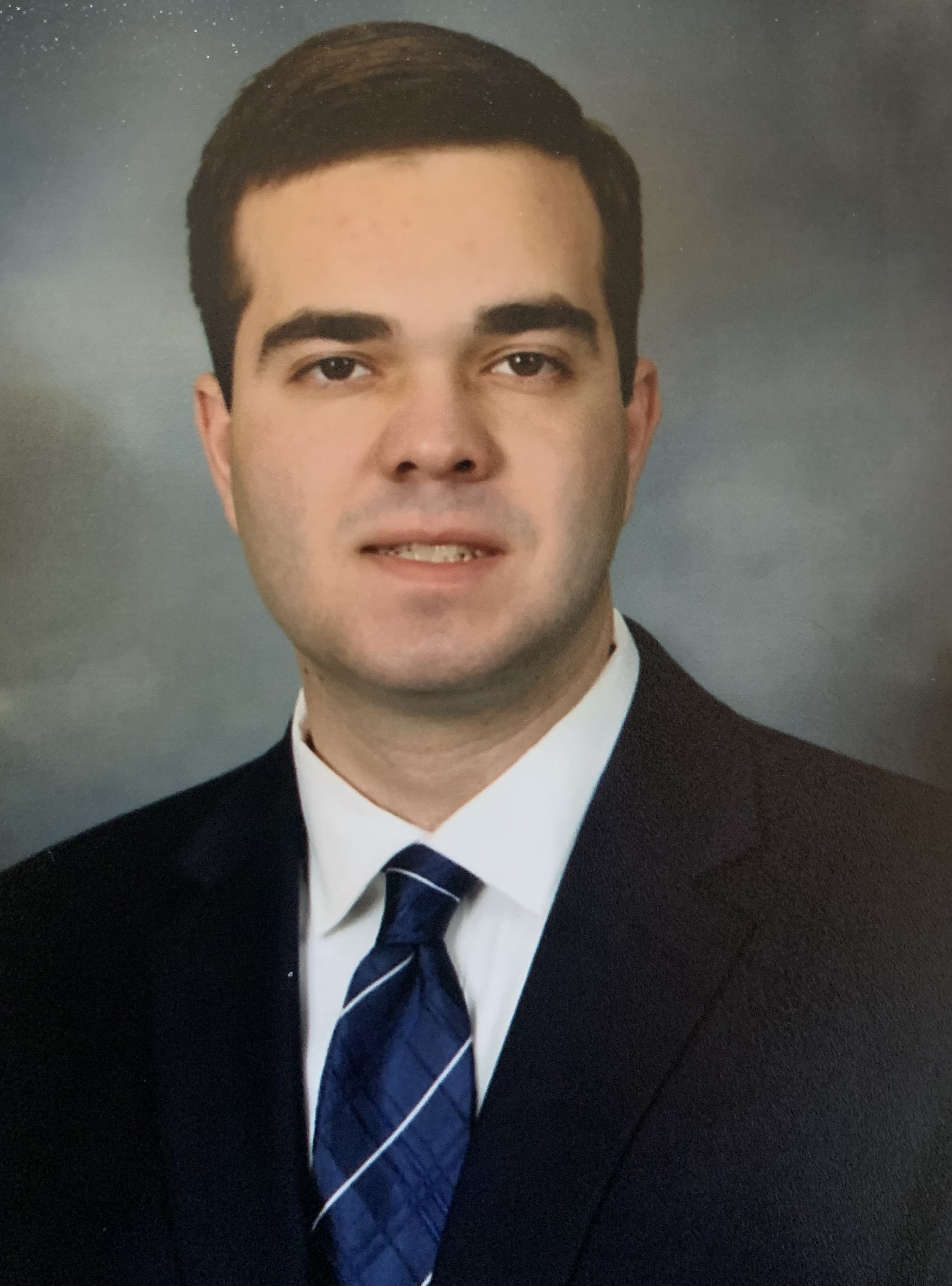}}]%
{Anthony D'Onofrio Jr.}  is a former graduate student from the Department of Computer and Information Science at Fordham University in New York City. He received his bachelor’s degree in Computer Science from Fordham University in 2022 and his master’s degree in May 2023. During his time at Fordham, he was a Fordham-IBM research intern and was inducted into Fordham University’s Chapter of Sigma Xi, the Scientific Research Honor Society, as an Associate Member for his work. His research interests focus on the fields of distributed systems, quantum systems, quantum deep learning, and software engineering.
\end{IEEEbiography}

\vskip -2.5\baselineskip plus -1fil

\begin{IEEEbiography}[{\includegraphics[width=1in,height=1.25in,clip,keepaspectratio]{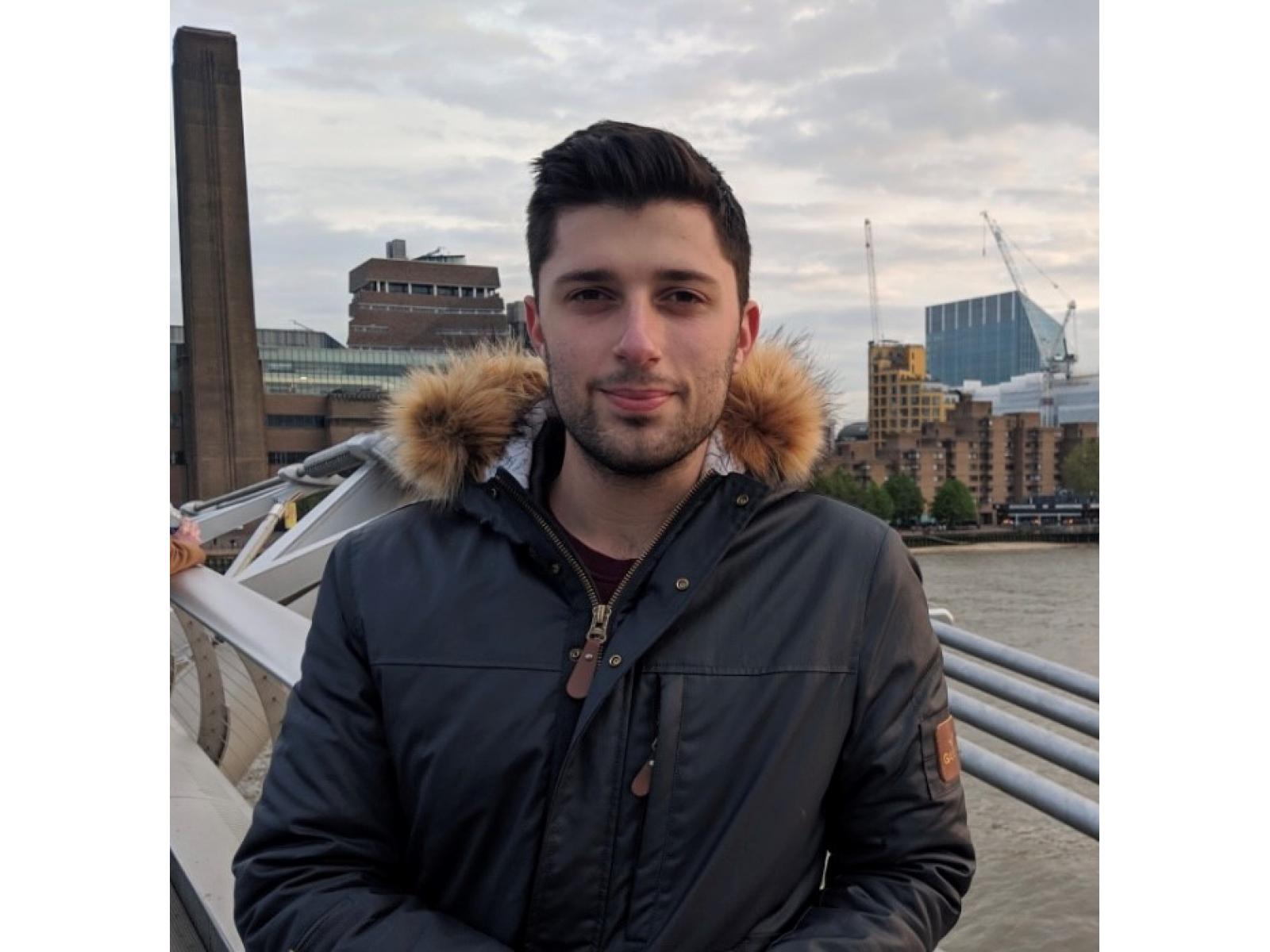}}]%
{Samuel Stein} is a Staff Scientist in the high-performance-computing (HPC) group of Pacific Northwest National Laboratory (PNNL) since December, 2022. He received his bachelors degree in Chemical Engineering from the University of Cape Town, South Africa, in 2018 and his masters in Data Science from Fordham University in 2020. His research has been focusing on Quantum Machine Learning, Quantum Error Mitigation, and Distributed Quantum Computing. More recently, his research has focused on heterogeneous quantum computing designs, and distributed quantum computing architectures.
\end{IEEEbiography}

\vskip -2.5\baselineskip plus -1fil

\begin{IEEEbiography}[{\includegraphics[width=1in,height=1.25in,clip,keepaspectratio]{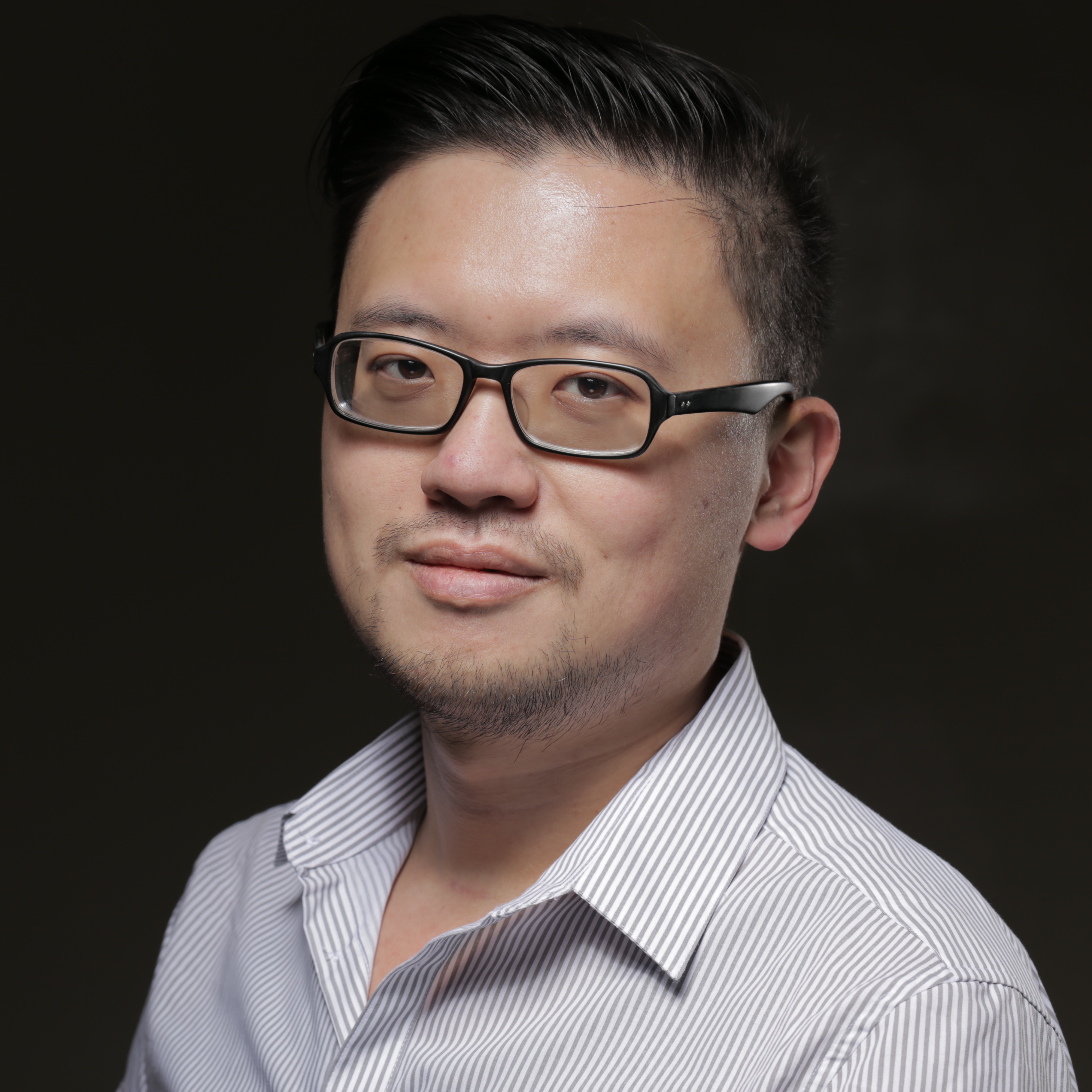}}]{Samuel Yen-Chi Chen} was an assistant computational scientist at Brookhaven National Laboratory, Upton, NY, USA. Dr. Chen received his Ph.D. degree in physics from National Taiwan University, Taipei, Taiwan, in 2020. He received the B.S. degree in physics and M.D. degree in medicine from National Taiwan University, Taipei, Taiwan, in 2016. His research focus on combining quantum computing and machine learning. He was the recipient of Theoretical High-Energy Physics Fellowship from Chen Cheng Foundation in 2014
and First Prize in the Software Competition (Research Category) from Xanadu Quantum Technologies in 2019. 
\end{IEEEbiography}
\vskip -2.5\baselineskip plus -1fil
\begin{IEEEbiography}[{\includegraphics[width=1in,height=1.1in,clip,keepaspectratio]{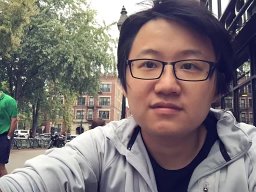}}]%
{Ang Li} is a senior computer scientist in the Physical and Computational Directorate of Pacific Northwest National Laboratory and affiliated Associate Professor at University of Washington, WA, USA. He received B.E. from Zhejiang University, China, and two PhD from the National University of Singapore and Eindhoven University of Technology, Netherlands. His research focuses on software-hardware codesign for scalable heterogeneous HPC and quantum computing.
\end{IEEEbiography}

\vskip -2.5\baselineskip plus -1fil

\begin{IEEEbiography}
[{\includegraphics[width=1in,height=1.25in,clip,keepaspectratio]{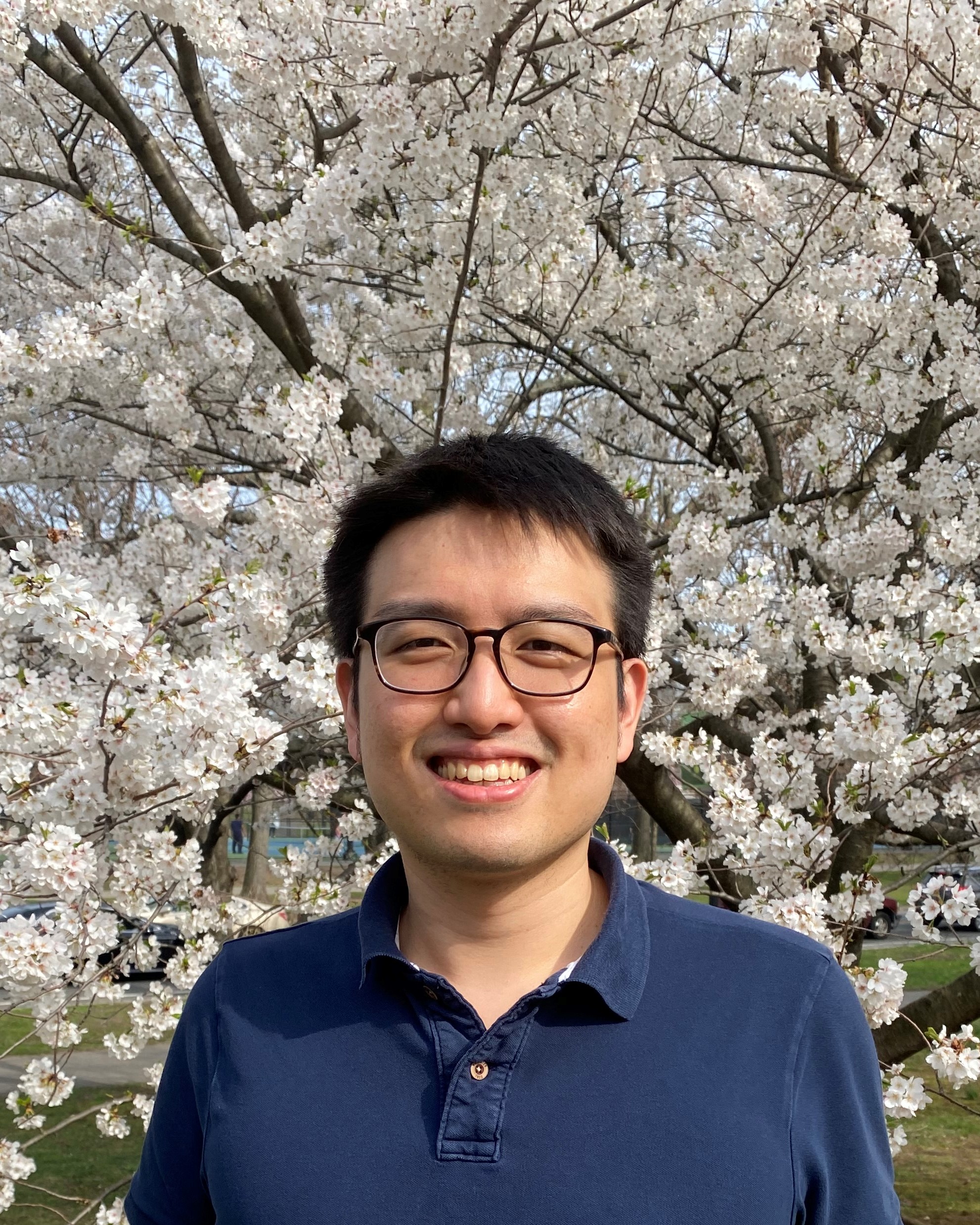}}]%
{Pin-Yu Chen} is a principal research staff member at IBM Research, Yorktown Heights, NY, USA. He is also the chief scientist of RPI-IBM AI Research Collaboration and PI of ongoing MIT-IBM Watson AI Lab projects. Dr. Chen received his Ph.D. degree in electrical engineering and computer science from the University of Michigan, Ann Arbor, USA, in 2016. Dr. Chen’s recent research focuses on adversarial machine learning and robustness of neural networks. His long-term research vision is to build trustworthy machine learning systems.  At IBM Research, he received the honor of IBM Master Inventor and several research accomplishment awards, including an IBM Master Inventor and IBM Corporate Technical Award in 2021. His research works contribute to IBM open-source libraries including Adversarial Robustness Toolbox (ART 360) and AI Explainability 360 (AIX 360). He has published more than 50 papers related to trustworthy machine learning at major AI and machine learning conferences.
He is a member  of IEEE and an associate editor of Transactions on Machine Learning Research.
\end{IEEEbiography}
\vskip -2.5\baselineskip plus -1fil

\begin{IEEEbiography}[{\includegraphics[width=1in,height=1.25in,clip,keepaspectratio]{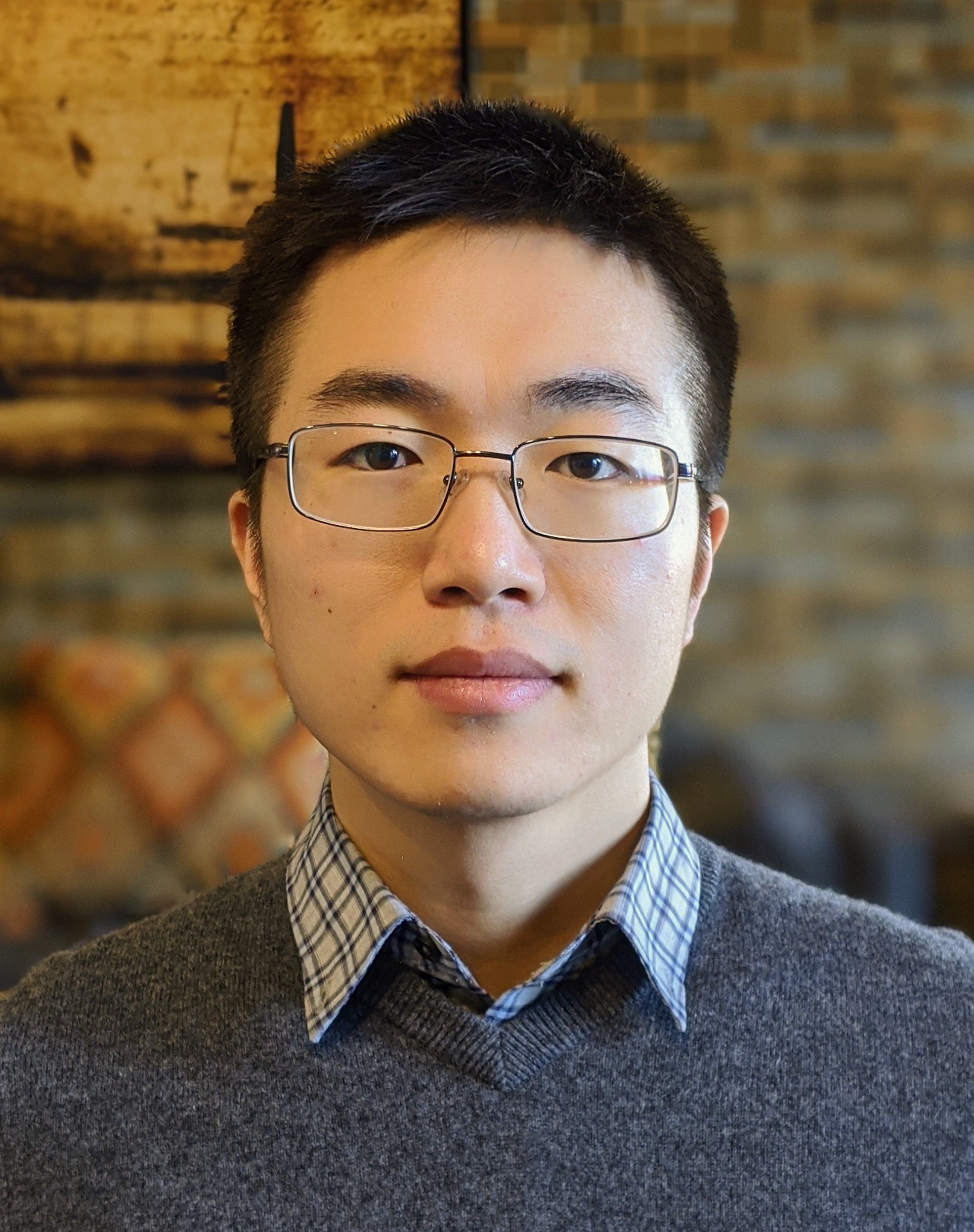}}]{Juntao Chen}
(S'15-M'21) received the Ph.D. degree in Electrical Engineering from New York University (NYU), Brooklyn, NY, in 2020, and the B.Eng. degree in Electrical Engineering and Automation with honor from Central South University, Changsha, China, in 2014. He is currently an assistant professor at the Department of Computer and Information Sciences and an affiliated faculty member with the Fordham Center of Cybersecurity, Fordham University, New York, USA. His research interests include cyber-physical security and resilience, quantum AI and its security, game and decision theory, network optimization and learning. He was a recipient of the Ernst Weber Fellowship, the Dante Youla Award, and the Alexander Hessel Award for the Best Ph.D. Dissertation in Electrical Engineering from NYU.
\end{IEEEbiography}
\vskip -2.5\baselineskip plus -1fil

\begin{IEEEbiography}[{\includegraphics[width=1in,height=1.25in,clip,keepaspectratio]{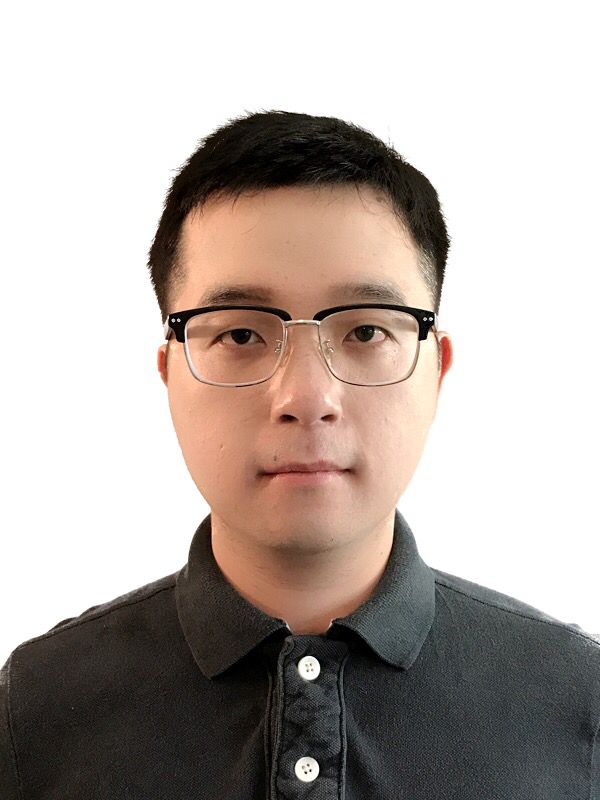}}]{Ying Mao} is an Associate Professor in the Department of Computer and Information Science at Fordham University in the New York City. He received his Ph.D. in Computer Science from the University of Massachusetts Boston in 2016. He was a Fordham-IBM research fellow. His research interests mainly focus on the fields of quantum systems, quantum deep learning, quantum-classical optimizations, quantum system virtualization, cloud resource management, data-intensive platforms and containerized applications.
\end{IEEEbiography}

\EOD
\end{document}

%% file: sources/intro.tex

\section{INTRODUCTION}  





Recent years have witnessed significant progress in machine learning and deep learning. 
Groundbreaking models and algorithms have significantly enhanced our capabilities to identify patterns and process data in areas such as computer vision, natural language processing, and finance. However, this accelerated development has led to an exponential increase in the computational power needed to execute increasingly sophisticated deep learning tasks.
As the era of Moore's Law comes to a close, however, the acceleration of computational demand is starting to surpass the growth in available computing power~\cite{thompson2020}. 
Consequently, this trend fuels the search for alternative computing approaches capable of managing the ever-growing computational needs. 


Quantum computing provides considerable potential in delivering the increased computational power essential to meet the expanding demands of deep learning. Classical computers employ binary bits, representing either 0 or 1, which constitute the current computing standard. In contrast, quantum computers use quantum bits (or qubits), which are probabilistic combinations of 0 and 1, achieved through quantum {\em superposition} and {\em entanglement}. As a result, the expected value of a qubit measurement can represent any number between 0 and 1. Therefore, a specific number of qubits can exhibit substantially greater representational power compared to an equivalent number of classical bits. 
In 1998, the first quantum computer capable of executing computations was developed ~\cite{chuang1998}. 
The IBM-Q Experience was introduced in 2016, granting developers access to state-of-the-art quantum resources \cite{ibmq}. 
In 2020, Google AI demonstrated that a 53-qubit quantum computer could complete a task in 200 seconds that would require a classical computer more than 10,000 years. This advantage of quantum computing over classical computing is frequently referred to as "quantum supremacy" ~\cite{arute2019}. 


Researchers inspired by the concept of quantum supremacy are actively exploring methods to convert classical algorithms into their quantum versions, aiming to achieve significant reductions in time complexity compared to classical counterparts. 
Quantum speed-ups have already been demonstrated for Shor's algorithm~\cite{shor1996} , which addresses prime factorization and discrete logarithms, and Grover's algorithm, which tackles database searches~\cite{grover1996}. Quantum computing can be applied to machine learning tasks by employing variational quantum circuits—quantum circuits with trainable parameters.
Specific areas within classical learning, such as Deep Learning and Support Vector Machines, could potentially benefit from quantum computing ~\cite{garg2020,beer2020}. Quantum speed-ups have been achieved for several algorithms, including expectation maximization solving ~\cite{kerenidis2019} (where the algorithm's speed has been increased to sub-linear time ~\cite{li2019}), Support Vector Machines ~\cite{ding2019}, and natural language processing ~\cite{panahi2019}.

However, in the noisy intermediate-scale quantum (NISQ) era, the qubits are  both limited in number and subject to noise. For instance, IBM-Q provides only 5-7 qubit machines to the public. Furthermore, as the qubit count increases, the computational complexity of the system grows exponentially \cite{kaye2007}, which leads to a higher overall noise level in a quantum machine. In the context of deep learning, an increased number of qubits may employ a greater number of gates, potentially augmenting circuit depth and noise interference. Consequently, it is crucial to efficiently and reliably utilize the representational power of qubits through effective encoding, making quantum algorithms more feasible on both current and NISQ quantum computers, while mitigating the surge in computational complexity as the number of qubits increases.
A potential solution to data encoding challenges involves performing classical pre-processing of the data for compression and/or feature extraction. One prevalent method for dimension reduction is Principal Component Analysis (PCA), as demonstrated in prior works~\cite{stein2021, stein2021hybrid, stein2021qugan, chen2021end, hur2022quantum}. However, PCA may not possess the representational power necessary to compress data accurately. More sophisticated methods, such as employing neural network layers, demand substantial pre-training and significantly increase the number of parameters requiring tuning. Therefore, there is a pressing need for efficient data compression techniques tailored to quantum machine learning.


In this work, we introduce a novel classical-quantum collaborative training architecture, which incorporates a classical tensor network (TN) into the feature extraction stage to facilitate dimensionality reduction. Specifically, the TN serves as a trainable module designed to capture high-level abstractions of the input data, the output of which is subsequently fed into a variational quantum circuit (VQC) for classification purposes. Furthermore, we employ a quantum-state-fidelity based cost function to train the model directly on qubits' states. Our proposed solution presents significant advantages over existing techniques, such as Principal Component Analysis (PCA), which lacks trainability, and conventional neural networks that require a considerable number of parameters to be optimized or pre-trained. The integration of our hybrid system enables more efficient data encoding, thereby enhancing the overall performance of the quantum machine learning pipeline. The main contributions are summarized as
follows.

\begin{itemize}
    \item We propose \sol, a quantum-classical collaborative training architecture. On the classical part, it employs tensor network layers for data pre-processing and preparation. In the quantum part, it utilizes a pre-processed dataset to build circuits with fewer qubits to reduce the overall qubit requirement and noise interference. 
    \item We introduce a quantum state fidelity based cost function. Instead of converting back to classical states, \sol~ train the model directly on quantum states aiming at accelerating the training process and improving performance. 
    \item We implement \sol~ with popular quantum toolkits, e.g., Qiskit and PennyLane, and compare it with state-of-the-art solutions in the literature, by up to 1.9x and 70.59\% less quantum resources. Additionally, we conduct proof-of-concept experiments on 14 different IBM-Q quantum machines.  
\end{itemize}





%% file: sources/related.tex
\section{RELATED WORK}  

Recent developments~\cite{houssein2022machine, massoli2022leap, cerezo2022challenges, chang2021fundamentals, ruan2022vacsen, donofrio2023distributed, ruan2023quantumeyes, ruan2023venus} in quantum computing show great potential to enhance current learning algorithms through utilization of the qubit, the unit of quantum information. 
In this field, quantum neural networks (QNN) have emerged as a promising research area in quantum machine learning~\cite{beer2020training, abbas2021power, xue2021quantum}. 
Due to the limited quantum resources available, most of the existing works focused on numerical  analysis or datasets with lower dimensionalities~\cite{chen2021end, liang2022variational, easom2022efficient}, such as MNIST~\cite{MNIST}.

Farhi et al.~\cite{farhi2018} introduced a QNN for binary classification, which utilizes quantum entanglement to enhance the model's computational power. In addition, quantum circuit learning~\cite{mitarai2018quantum, ostaszewski2021reinforcement} developed a  quantum-classical hybrid algorithm. They employed an iterative optimization of the parameters to circumvent the high-depth circuit. 
Moreover, Stokes et al. \cite{Stokes_2020} presented a novel method for gradient descent using quantum circuits, enabling the optimization of variational quantum circuits in a manner analogous to classical neural networks. 
However, these solutions focused on theoritical analysis and only numerical experiments were provided.


In NISQ era, QCNN~\cite{Cong_2019} suggests a design for a quantum convolutional neural network that uses $O(log(N))$ trainable parameters for $N$ dimensional inputs 
and can be realized on near-term quantum computers.
Additionally, QuCNN~\cite{stein2022qucnn} employs an entanglement based backpropagation for NISQ machines. 
Jiang et al. \cite{Jiang_2021} proposed a co-design framework named QuantumFlow, which features quantum-friendly neural networks, a mapping tool to generate quantum circuits, and an execution engine. However, QuantumFlow requires local training of the network prior to mapping to quantum circuits, which leads to sensitivity to noise when implemented on real quantum computers as opposed to simulations. 

Expanding upon the use of quantum operations to perform distance measurements, Stein et. al proposed the QuClassi system: a hybrid quantum-classical system with a quantum-state-fidelity based loss function \cite{stein2021, stein2021hybrid}. QuClassi  was able to provide improvements in accuracy compared to other contemporary quantum-based solutions such as TensorFlow Quantum~\cite{broughton2020tensorflow}
and QuantumFlow. The QuClassi system demonstrated success in both binary and multi-class classification. It used Principal Component Analysis (PCA) to compress dataset classically. 
However, PCA fails to fully utilize the classical resources by providing trainable layers. 
TN-VQC~\cite{chen2020} proposed the use of tensor networks for feature extraction and data compression to achieve higher classification accuracy for variational quantum circuits. Tensor networks do provide the advantage of having fewer parameters compared to neural networks while still providing some trainability unlike PCA. TN-VQC employed a circuit architecture involving CNOT gates rather than CSWAP gates like QuClassi.

This paper proposes \sol, a hybrid quantum-classical architecture for deep neural networks. Comparing with existing literature, it utilizes a quantum-state fidelity based cost function to train the quantum section directly on qubits' states. Additionally, tensor networks are employed to fully exploit classical resources to compensate for the limitations (e.g., low qubit count and noises) of quantum resources.  Through a collaborative training process, \sol~ is able to outperform state-of-the-arts.




%% file: sources/motivation.tex
 \section{BACKGROUND}  

In this section, we present the background that is necessary for designing our solution.
\subsection{Quantum Computing Basics}
\subsubsection{A Qubit and its superposition}
\label{super}
Classical computing uses bits that are binary in nature and measure either 0 or 1. Quantum computing uses quantum bits or qubits. Qubits, unlike classical bits, are a probabilistic mixture of 0 and 1. This mixture of 0 and 1 is known as a superposition. Upon measurement, the qubit in superposition will collapse to either a value of 0 or 1. Quantum circuits are often run many times, using the results to get a probability distribution for the circuit results. Calculations are performed by manipulating the probability distributions of qubits. 0 and 1 can be represented in vector notation as seen in Equation \ref{eq:base_equations-2}. 

Quantum systems are often described using $\langle bra|$ $|ket \rangle$ notation, where $\langle bra|$ and $|ket\rangle$ represent horizontal and vertical quantum state vectors, respectively. Because a qubit is a mixture of 0 and 1, qubit states are described mathematically as a linear combination of $|0\rangle$  and $|1\rangle$ as seen in Equation \ref{eq:base_equations-2} and \ref{eq:q_state_equation}.

\begin{equation}
|0\rangle=\left[\begin{array}{l}
1 \\
0
\end{array}\right]\text{ , }|1\rangle=\left[\begin{array}{l}
0 \\
1
\end{array}\right]\text{ , }|\Psi\rangle=\left[\begin{array}{l}
\alpha \\
\beta
\end{array}\right]
\label{eq:base_equations-2}
\end{equation}

\begin{equation}
|\Psi\rangle=\alpha|0\rangle + \beta|1\rangle
\label{eq:q_state_equation}
\end{equation}

 This linear combination of qubit states is referred to as a qubit's statevector. $|0\rangle$  and $|1\rangle$ are orthonormal vectors in an eigenspace. In Equation \ref{eq:q_state_equation}, $|\Psi\rangle$ represents the qubit state, a probabilistic combination of $|0\rangle$  and $|1\rangle$.

The tensor product of qubit states can be used to describe the quantum states of multiple qubits. The tensor product between the qubits shown in Equations \ref{eq:q_state_equation} and \ref{eq:q_state_equation2} can be described using Equation \ref{eq:q_state_product}.

\begin{equation}
|\Phi\rangle=\gamma|0\rangle + \omega|1\rangle
\label{eq:q_state_equation2}
\end{equation}

\begin{equation}
|\Psi\Phi\rangle=|\Psi\rangle \otimes |\Phi\rangle =  \gamma\alpha|00\rangle + \omega\alpha|01\rangle + 
\gamma\beta|10\rangle + \omega\beta|11\rangle
\label{eq:q_state_product}
\end{equation}

$|0\rangle$  and $|1\rangle$ represent opposite points of the sphere on the z axis. Measurements of qubit states can be taken with respect to any basis, but convention typically dictates that measurements are taken against the z-axis. However, the x-axis, y-axis, or any pair of opposite points on the sphere could potentially be used as a basis of measurement.
Quantum states are responsible for encoding data, and to perform operations on quantum states quantum gates are used. Quantum gates apply a transformation over a quantum state into some new quantum state.

\subsection{Quantum Gates}
Similar to classical data which is manipulated and encoded using gates, quantum data is manipulated and encoded using quantum gates. Quantum gates can either perform a rotation about an axis or perform an operation on a qubit based on the value of another qubit. These are referred to as rotation gates and controlled gates respectively.

\subsubsection{Single-Qubit Gates}

A common type of single-qubit operations are the rotation gates. These gates perform qubit rotations by parameterized amounts. The generalized single-rotation gate $R$ is shown in matrix form in Equation \ref{eq:rotation_general}.

\begin{equation}
R(\theta,\phi)=\left[\begin{array}{cc}
\cos \frac{\theta}{2} & -ie^{-i\phi}\sin \frac{\theta}{2} \\
-ie^{-i\phi}\sin \frac{\theta}{2} & \cos \frac{\theta}{2}
\end{array}\right]
\label{eq:rotation_general}
\end{equation}

Three commonly-used special cases of this gate are the $R_{X}$, $R_{Y}$, and $R_{Z}$ gates. These gates represent rotations in the x, y, and z plane and are expressed in Equations \ref{eq:rx}, \ref{eq:ry}, and \ref{eq:rz}. $R_{X}$ and $R_{Y}$ can be thought of as special cases of the $R$ gate in which $\phi = 0$ and $\phi = \frac{\pi}{2}$ respectively. Therefore, $R_{X}(\theta)$ is a rotation about the x-axis by angle $\theta$ and $R_{Y}(\theta)$ is a rotation about the y-axis by angle $\theta$. The derivation of $R_{Z}$ from the general rotation gate is less straightforward and thus is not included here.

\begin{equation}
R_{X}(\theta)=\left[\begin{array}{cc}
\cos \frac{\theta}{2} & -i\sin \frac{\theta}{2} \\
-i\sin \frac{\theta}{2} & \cos \frac{\theta}{2}
\end{array}\right]=R(\theta,0)
\label{eq:rx}
\end{equation}

\begin{equation}
R_{Y}(\theta)=\left[\begin{array}{cc}
\cos \frac{\theta}{2} & -\sin \frac{\theta}{2} \\
\sin \frac{\theta}{2} & \cos \frac{\theta}{2}
\end{array}\right]=R(\theta,\frac{\pi}{2})
\label{eq:ry}
\end{equation}

\begin{equation}
R_{Z}(\theta)=\left[\begin{array}{cc}
e^{\frac{-i\theta}{2}} & 0 \\
0&e^{\frac{-i\theta}{2}}
\end{array}\right]
\label{eq:rz}
\end{equation}

\subsubsection{Hadamard Gate}

A fundamental gate of quantum computation is the Hadamard gate. It is a single-qubit gate puts a qubit into superposition as described in Section~\ref{super}. It can be expressed in matrix shown in equation~\ref{eq:hgate}. The $\frac{1}{\sqrt{2}}$ coefficient is due to the fact that the sum of the squares of the state amplitudes must add to 1, so each state has a probability of $\frac{1}{2}$ and an amplitude of $\frac{1}{\sqrt{2}}$.

\begin{equation}
H=\frac{1}{\sqrt{2}}\left[\begin{array}{cc}
1 & 1 \\
1 & -1
\end{array}\right]
\label{eq:hgate}
\end{equation}

\subsubsection{Two-Qubit Gates}

There are also operations that function as two-qubit rotations which perform an equal rotation on two qubits. These gates are described in Equations \ref{eq:rxx}, \ref{eq:ryy}, and \ref{eq:rzz}. Note that these gates are expressed as 4x4 matrices while the single-qubit gates were 2x2 matrices. This is because for a two-qubit gate, each individual qubit has two possible measurements, yielding four possible results ($|00\rangle$,$|01\rangle$,$|10\rangle$,$|11\rangle$) rather than two as seen previously for the single-qubit gates.

\begin{small}
\begin{equation}
R_{XX}(\theta)=\left[\begin{array}{cccc}
\cos \frac{\theta}{2} & 0 & 0 & -i\sin \frac{\theta}{2}\\
0 & \cos \frac{\theta}{2} & -i\sin \frac{\theta}{2} & 0 \\
0 & -i\sin \frac{\theta}{2} & \cos \frac{\theta}{2} & 0 \\
-i\sin \frac{\theta}{2} & 0 & 0 & \cos \frac{\theta}{2}
\end{array}\right]
\label{eq:rxx}
\end{equation}
\end{small}

\begin{small}
\begin{equation}
R_{YY}(\theta)=\left[\begin{array}{cccc}
\cos \frac{\theta}{2} & 0 & 0 & i\sin \frac{\theta}{2}\\
0 & \cos \frac{\theta}{2} & -i\sin \frac{\theta}{2} & 0 \\
0 & -i\sin \frac{\theta}{2} & \cos \frac{\theta}{2} & 0 \\
i\sin \frac{\theta}{2} & 0 & 0 & \cos \frac{\theta}{2}
\end{array}\right]
\label{eq:ryy}
\end{equation}
\end{small}

\begin{small}
\begin{equation}
R_{ZZ}(\theta)=\left[\begin{array}{cccc}
e^{-i\frac{\theta}{2}} & 0 & 0 & 0\\
0 & e^{-i\frac{\theta}{2}} & 0 & 0 \\
0 & 0 & e^{-i\frac{\theta}{2}} & 0 \\
0 & 0 & 0 & e^{-i\frac{\theta}{2}}
\end{array}\right]
\label{eq:rzz}
\end{equation}
\end{small}

\subsubsection{Controlled Gates}

There are also two-qubit gates which utilize a control qubit and a target qubit. These gates, known as controlled gates, perform an operation on a target qubit depending on the value of the control qubit. 

{\bf \noindent CNOT Gate}
The CNOT gate is an example of a two-qubit gate used in quantum computing. The CNOT gate flips the value of the target qubit if the control qubit is measured as 1 and does nothing otherwise. The CNOT gate can be seen represented in matrix form below. 

\begin{equation}
CNOT=\left[\begin{array}{cccc}
1 & 0 & 0 & 0 \\
0 & 0 & 0 & 1 \\
0 & 0 & 1 & 0 \\
0 & 1 & 0 & 0
\end{array}\right]
\label{eq:squ-3}
\end{equation}

\begin{figure}[ht]			
	\centering
	\includegraphics[width=0.3\linewidth]{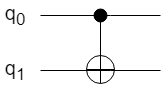}
    \caption{CNOT Gate Circuit Notation}
    \label{fig:cnot_gate}
\end{figure}

Fig.\ref{fig:cnot_gate} depicts the circuit notation for the CNOT gate. $q_{0}$ is the control qubit and $q_{1}$ is the target qubit.

{\bf \noindent Controlled Rotation Gates}
Equations \ref{eq:crx}, \ref{eq:cry} and \ref{eq:crz} are controlled rotation gates in matrix notation. Controlled rotation gates are similar to the CNOT gate but apply a rotation when the control qubit measures 1 instead of flipping the state. This allows for variable levels of entanglement between qubits.

\begin{equation} 
\centering
CR_X(\theta)=\left[\begin{array}{cccc} 
1 & 0 & 0 & 0 \\
0 & 1 & 0 & 0 \\
0 & 0 & \cos\frac{\theta}{2} & -\sin\frac{\theta}{2} \\
0 & 0 & -\sin\frac{\theta}{2} & \cos\frac{\theta}{2} 
\end{array}\right]
\label{eq:crx}
\end{equation}

\begin{equation} 
\centering
CR_Y(\theta)=\left[\begin{array}{cccc} 
1 & 0 & 0 & 0 \\
0 & 1 & 0 & 0 \\
0 & 0 & \cos\frac{\theta}{2} & -\sin\frac{\theta}{2} \\
0 & 0 & \sin\frac{\theta}{2} & \cos\frac{\theta}{2} 
\end{array}\right]
\label{eq:cry}
\end{equation}

\begin{equation} 
\centering
CR_Z(\theta)=\left[\begin{array}{cccc} 
1 & 0 & 0 & 0 \\
0 & 1 & 0 & 0 \\
0 & 0 & e^{\frac{i\theta}{2}} & 0 \\
0 & 0 & 0 & e^{\frac{i\theta}{2}} 
\end{array}\right]
\label{eq:crz}
\end{equation}

\subsection{Controlled Swap Gate}
Another type of controlled gate is the controlled SWAP gate. The SWAP gate measures the difference between two quantum states and outputs the result to an ancilla qubit. Therefore, this gate is a three-qubit gate. The SWAP test output values range from 0.5 to 1. Maximally different (orthogonal) states will measure 1 with 50\% probability while identical states will measure 1 with 100\% probability. The SWAP test gate can be used to measure quantum state fidelity. The controlled swap gate is described in Equations \ref{eq:cswap_eq} and \ref{eq:cswap_matrix}.

\begin{equation} 
\centering
CSWAP(q_{0},q_{1},q_{2})=|0\rangle \langle0|\otimes I \otimes I + |1\rangle \langle1| \otimes SWAP
\label{eq:cswap_eq}
\end{equation}

\begin{small}
\begin{equation} 
\centering
CSWAP(q_{0},q_{1},q_{2})=\left[\begin{array}{cccccccc} 
1 & 0 & 0 & 0 & 0 & 0 & 0 & 0\\
0 & 1 & 0 & 0 & 0 & 0 & 0 & 0\\
0 & 0 & 1 & 0 & 0 & 0 & 0 & 0\\
0 & 0 & 0 & 0 & 0 & 1 & 0 & 0\\
0 & 0 & 0 & 0 & 1 & 0 & 0 & 0\\
0 & 0 & 0 & 1 & 0 & 0 & 0 & 0\\
0 & 0 & 0 & 0 & 0 & 0 & 1 & 0\\
0 & 0 & 0 & 0 & 0 & 0 & 0 & 1\\
\end{array}\right]
\label{eq:cswap_matrix}
\end{equation}
\end{small}

\begin{figure}[ht]			
	\centering
	\includegraphics[width=0.5\linewidth]{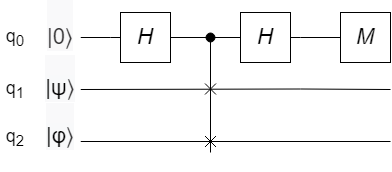}
    \caption{Swap Test Quantum Circuit}
    \label{fig:swap_test}
\end{figure}

Figure \ref{fig:swap_test} depicts a swap test being performed. The ancilla qubit, $q_0$, is placed is superposition using a Hadamard gate. Then a swap test is performed between $q_1$ and $q_2$ and measured onto $q_0$. Another Hadamard gate is performed on the ancilla qubit. Finally, the ancilla qubit is then measured onto a classical bit to obtain the result.

One advantage of the CSWAP gate is that it only requires the measurement of the ancilla qubit. When qubits are measured directly, their states collapse and the superposition is lost. The Swap test allows the superposition of the other qubits to be maintained by measuring the quantum state fidelity through the ancilla qubit instead of measuring the qubits directly. Therefore, minimal information is lost through measurement.

\subsection{Quantum Entanglement}
A key principle of quantum computing is quantum entanglement. A qubit’s state is said to be entangled when its measurement is dependent on the measurement of another qubit. This dependence allows information to be transferred between qubits, even if they are not physically close together (a phenomena sometimes referred to as "action at a distance"). When one entangled qubit is measured, the other entangled qubit's state also collapses. For example, if two qubits are entangled using the CNOT gate, after the state of one qubit is measured, the state of the second entangled qubit can be predicted with absolute certainty. Quantum entanglement is a key component of the quantum advantage over classical computing, as it is a property of quantum computing with no classical equivalent.

%% file: sources/system.tex
\section{SYSTEM DESIGN}  

\begin{figure*}[ht]			
	\centering
	\includegraphics[width=0.90\linewidth]{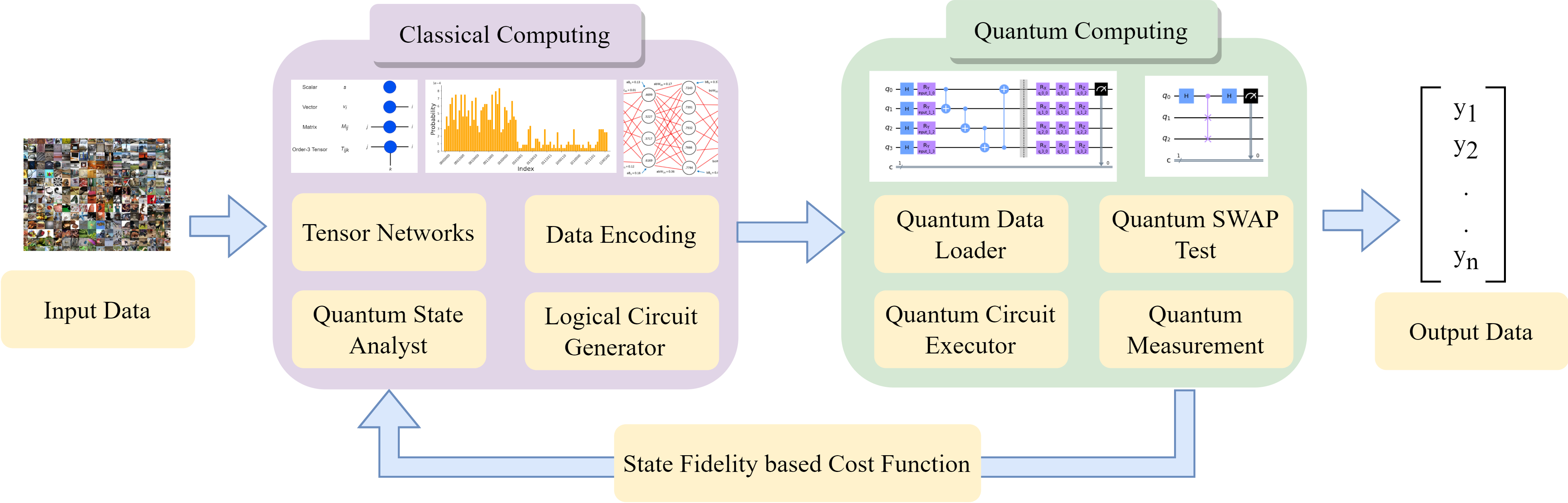}
    \caption{\sol: A Quantum-Classical Collaborative Training Architecture}
    \label{fig:System}
\end{figure*}

Our architecture employs a feedback loop between classical computers and quantum computers, as illustrated in Fig.~\ref{fig:System}. Initially, the is fed into tensor networks with a layers of trainable parameters and output the data in a pre-configured dimension. 
The data is then converted from classical data into quantum data through a quantum data encoding method, as outlined in Section~\ref{encode}. This results in a quantum data set represented by quantum state preparation parameters. For each predictable class in the data set, a quantum state is initialized with the same qubit count as the number of qubits in the classical quantum data set, due to the constraints of the SWAP test. The quantum states, along with quantum classical data, are then used to generate a logical quantum circuit and sent to a quantum computer for further processing.

This initialization of state is the core architecture to \sol. In this, a quantum circuit of a certain number of layers representing a quantum deep neural network (detailed in Section \ref{layers}) is prepared with randomly initialized parameters containing a certain number of qubits. The produced quantum state of this circuit is to be SWAP tested against the quantum data point, which is fed back to the classical computer and analyzed with quantum state fidelity based cost function (described in Section \ref{cost}), forming the overall collaborative quantum-classical deep learning architecture of \sol. 

The quantum computer calculates the quantum fidelity from one ancilla qubit which is used to calculate model loss, and sends this metric back to the classical computer. The classical computer uses this information to update the learn-able parameters in attempts to minimize the cost function. This procedure of loading quantum states, measuring state fidelity, updating states to minimize cost is iterated upon until the desired convergence or sufficient epochs have been completed.


\subsection{Data Encoding on Qubits}
\label{encode}
When evaluating quantum machine learning architectures on classical datasets, it is crucial to have a method for translating classical data into quantum states. One question that arises is how to represent a classical dataset in a quantum setting. Our architecture utilizes the expectation of a qubit to translate traditional numerical data points. To achieve this, data ${x_1,x_2,...,x_n}$ of dimension ${d}$ can be mapped onto a quantum setting by normalizing each dimension ${d_i}$ to fall within the range of ${0}$ to ${1}$. This is because a qubit's expectation can only take on values within this range. In contrast to classical computing, which requires a string of bits to represent the same number, encoding a single dimension data point only requires one qubit. To translate the traditional value $x_i$ into some quantum state, we perform a rotation around the Y axis parameterized by the following equation: 
\begin{equation}
    RY(\theta_{x_i}) = 2sin^{-1}(\sqrt{x_i})
\end{equation}
The $RY(\theta_{x_i})$ operation results in the expectation of a qubit being measured against the Z axis, corresponding to the $x_i$ value from the classical data that the qubit encodes. Building upon this concept, we can encode the second dimension of data across the X-Y plane. To achieve this, we employ two parameterized rotations on one qubit initialized in state $|0\rangle$ to prepare classical data in the quantum setting. To encode a data point, we apply the necessary rotations across ${\frac{d}{2}}$ qubits, with each rotation parameterized by the normalized value of that data point's corresponding dimension.
It is worth noting that the encoding of 2-dimensional data onto a single qubit may pose challenges for extreme values of $x$. However, we explore the dual dimensional encoding as a possible method of reducing high qubit counts and evaluate the performance when each dimension of data is encoded into one respective qubit solely through a RY Gate. This approach is validated by the fact that we never measure any of our qubits, but only their quantum fidelity through the SWAP test. As a result, we can bypass the superposition-collapsing issue inherent in this approach.We encode the second dimension of data on the same qubit through the following rotation:

\begin{equation}
    RZ(\theta_{x_{i+1}}) = 2sin^{-1}(\sqrt{x_i})
\end{equation}

When dealing with a limited number of qubits, methods that can reduce the number required are highly valuable. Unlike classical computers, which utilize formats such as integers and floats, classical data encoding in quantum states does not have a tried and tested method. Therefore, our approach may be subject to criticism. Nevertheless, our approach has been tested and proven to be a viable solution to the problem at hand.
Additionally, having knowledge of both the qubit's expectation across the Y and Z domains enables the reconstruction of classical data. Various methods for classical-to-quantum data encoding exist, ranging from encoding $2^n$ classical data points across $n$ qubits using state-vector encoding to encoding classical data into a binary representation on quantum states by translating a vector of binary values onto qubits. The former method is highly susceptible to noise, whereas the latter loses significant information in the process but is less susceptible to noise and exponential-sampling problems. Exponential data-encoding methods also exist and can be integrated into \sol~ since it does not directly perform quantum state tomography, making the data encoding section scalable.

The \sol~ quantum circuits consist of $n+1$ qubits, with $n$ representing the dimension of the input data. The input data is encoded on $n/2$ qubits, while trainable parameters are applied to the remaining $n/2$ qubits. Additionally, there is one ancilla qubit used for swap test measurements.

\subsection{Quantum Layers}
\label{layers}
Similar to classical artificial neural networks, quantum circuits can also be thought of as having layers. For a quantum circuit, these layers would be comprised of quantum gates.

In \sol, we define three quantum layer styles: single-qubit unitary, dual-qubit unitary, and controlled-qubit unitary. Each of these layer styles comprises rotations that serve as the trainable parameters in our quantum machine learning model. Defining these three types of layers enables system design at a higher level than individual gates.

\begin{figure}[ht]			
	\centering
	\includegraphics[scale=0.4]{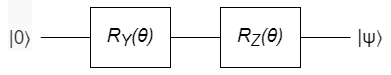}
    \caption{Single Qubit Unitary}
    \label{fig:single_qubit_unitary}
\end{figure}

{\noindent \bf Single-Qubit Unitary}
A single-qubit unitary layer involves single-qubit rotations around the y-axis and z-axis ($R_Y$ and $R_Z$). This allows for total manipulation of a qubit's quantum state. A single-qubit unitary layer is depicted in Figure \ref{fig:single_qubit_unitary}.

\begin{figure}[ht]
\begin{minipage}{0.45\columnwidth}
\centering
\includegraphics[width=1\columnwidth]{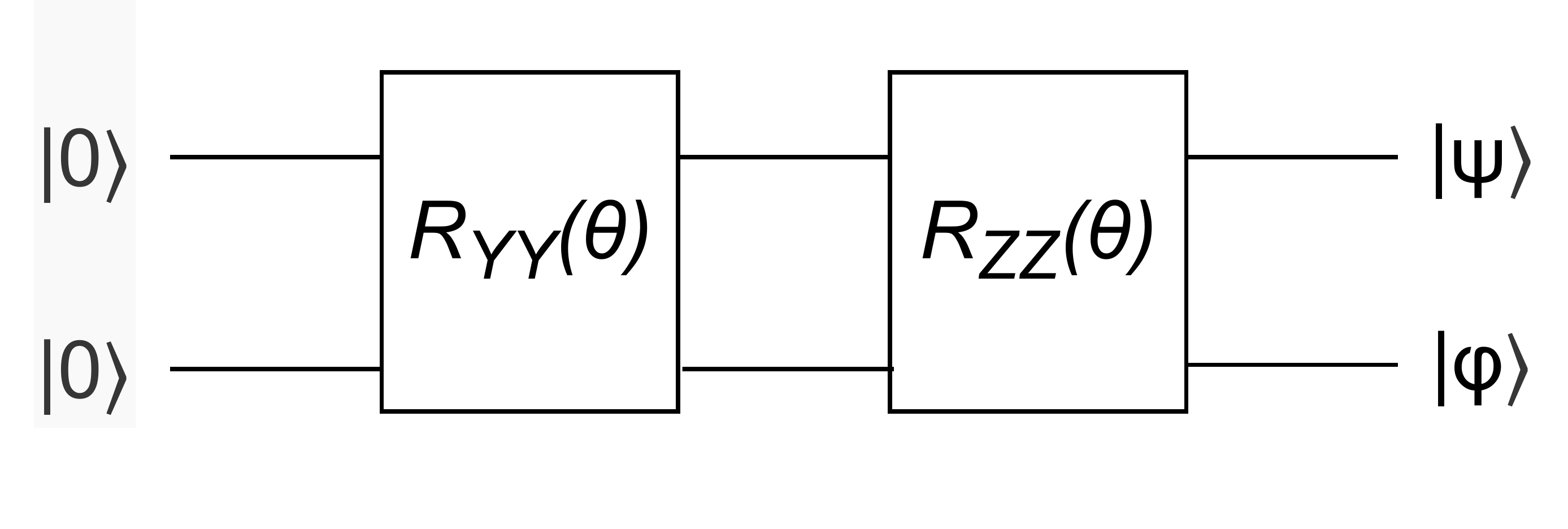}
\caption{Dual Qubit}
      \label{fig:dual_qubit_unitary}
\end{minipage}
\hfill
\begin{minipage}{0.45\columnwidth}
\centering
         \includegraphics[width=1\columnwidth]{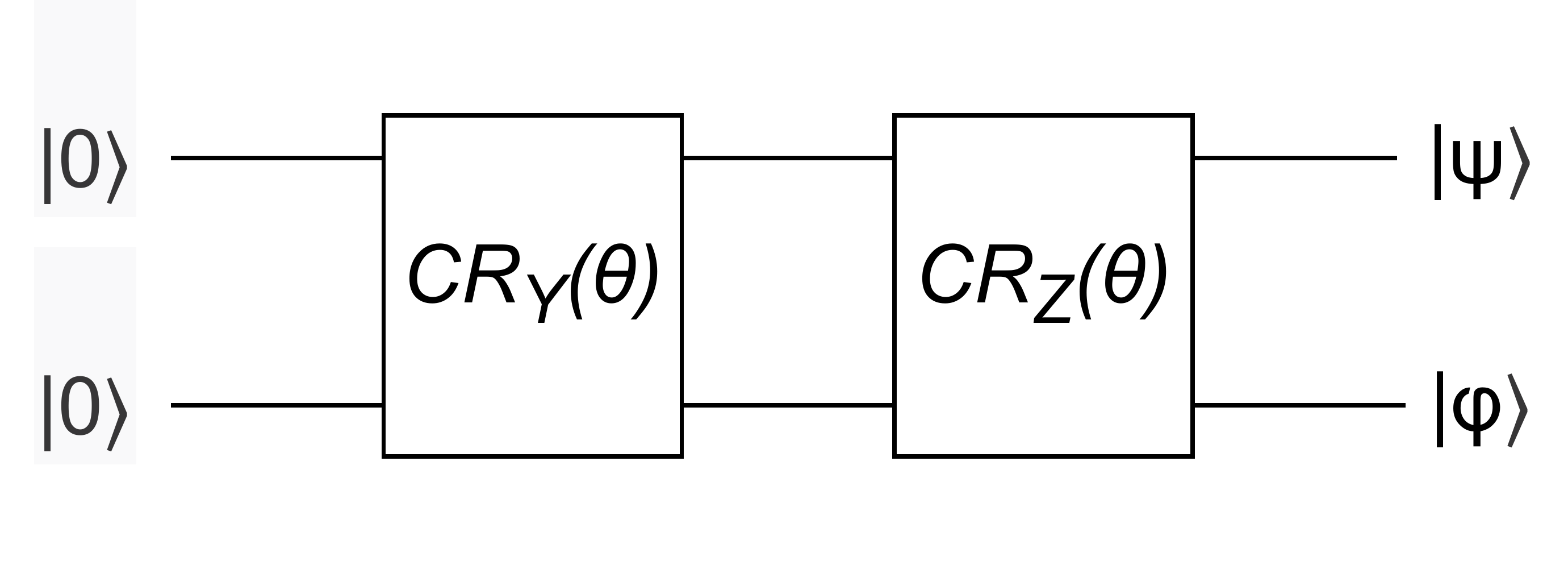}
\caption{Entanglement}
      \label{fig:controlled_qubit_unitary}
\end{minipage}      
\end{figure}

{\noindent \bf Dual-Qubit Unitary}
A dual-qubit untary layer involves dual-qubit rotations around the y and z axis ($R_{YY}$ and $R_{ZZ}$). The same y rotation and z rotation are applied to both qubits involved. A dual-qubit unitary layer is depicted in Figure \ref{fig:dual_qubit_unitary}.

{\noindent \bf Entanglement-based Unitary}
A controlled-qubit unitary utilizes controlled rotation gates ($CR_Y$ and $CR_Z$) to entangle qubits.The use of these gates allows the level of entanglement between qubits to be trainable. In Figure \ref{fig:controlled_qubit_unitary}, the top row is the control qubit and the bottom row is the target qubit.

\begin{figure}[ht]			
	\centering
	\includegraphics[width=1\linewidth]{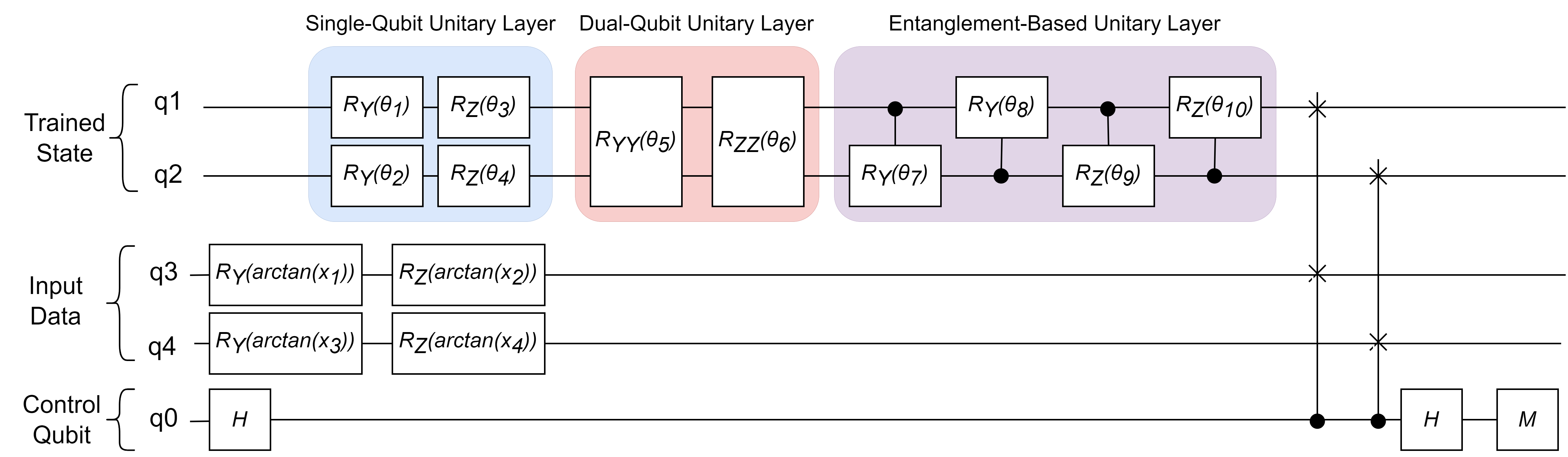}
    \caption{\sol~ with 3-layers and 5-qubits setting}
    \label{fig:quclassi-sde}
\end{figure}

The layers can be combine linearly to composite a multi-layer mode. For example, as seen in Figure \ref{fig:quclassi-sde}, the circuit features three layer types: single-qubit unitary, dual-qubit, unitary, and controlled-qubit unitary.

\subsection{Parameter Shift}

Backpropagation is a necessary step for training any deep neural network. Gradients for the parameters of quantum circuits cannot be calculated by the same methods used in classical backpropagation. Therefore, the gradients of the parameters are calculated using parameter shift shown in Equation \ref{eq:param_shift}.


\begin{equation}
\nabla_{\theta}f\left(\theta \right)=0.5*[f\left(\theta + s \right) - f\left(\theta - s\right)]
\label{eq:param_shift}
\end{equation}

With the parameter shift rule, the quantum circuit can be viewed as a black box and the gradient is calculated by obtaining circuit results when the parameter is increased or decreased by a shift $s$. The difference in results can be used to obtain a gradient for the parameter. 

\subsection{State fidelity based cost Function}
\label{cost}
When training a neural network to accomplish a task, an explicit description of system improvement goal needs to be established - i.e the cost function. The quantum machine learning cost function landscape can be slightly ambiguous compared to classical machine learning, as we could be manipulating the expected values of each qubit in some way. However, even this is ambiguous - the direction being measured in heavily affects the expectation value and or  what our iteration count would be for measuring expectation, with lower iterations leading to increasingly noisy outputs. Within our system, we make use of the SWAP test to parse quantum state fidelity to an appropriate cost function. One of the benefits of the SWAP test is that we only need to measure one ancilla qubit. In the case of binary classification, each data point is represented in a quantum state represented by $|\phi\rangle$, which is used to train the quantum state prepared by our DL model $|\omega\rangle$ such that the state of $|\omega\rangle$ minimizes some cost function. The classical cross-entropy cost function outlined in Equation \ref{eq:cross_ent} is an appropriate measure for state fidelity, as we want the fidelity returned to be maximized in the case of Class=1, and minimized otherwise.
\begin{equation}
\label{eq:cost_func}
    min(Cost(\theta_d,X) = \frac{1}{n}\sum_{i=1}^{n} SWAP(|\phi_{X(i)}\rangle,|\omega\rangle)
\end{equation}
\begin{equation}
\label{eq:cross_ent}
    Cost = -y log(p) - (1-y) log(1-p)
\end{equation}
Where $\theta_d$ is a collection of parameters defining a circuit, $x$ is the data set, $\phi_{x(i)}$ is the quantum state representation of data point ${i}$, and ${\omega}$ is the state being trained to minimize the function in Equation \ref{eq:cost_func} and \ref{eq:cross_ent}. \\
Optimization of the parameters $\theta_d$ requires us to perform gradient descent on our cost function. We make use of the following modified parameterized quantum gate differentiation formula outlined in Equation \ref{eq:diff}.
\begin{equation}
    \label{eq:diff}
    \frac{\delta Cost}{\delta\theta_i} = \frac{1}{2}(f(\theta_i + \frac{\pi}{2\sqrt{\epsilon}}) - f(\theta_i - \frac{\pi}{2\sqrt{\epsilon}}))
\end{equation}

Where in Equation \ref{eq:diff} $\theta_i$ is a parameter, Cost is the cost function, and $\epsilon$ is the epoch number of training the circuit. Our addition of the $\epsilon$ is targeted at allowing for a change in search-breadth of the cost landscape, shrinking constantly ensuring a local-minima is found.

The gradients of quantum parameters can also be determined using numerical methods. Equation \ref{eq:finite_diff} show a formula to numerically determine the gradients of quantum parameters. However, numerical methods can run into issues due to the noise an error associated with current quantum computers. Therefore, the gradients calculated may be inaccurate and lead to inefficiency in training \cite{bergholm2018}.

\begin{equation}
\nabla_{\theta}f\left(\theta \right)=\frac{f\left(\theta + s \right) - f\left(\theta - s\right)}{2s}
\label{eq:finite_diff}
\end{equation}

\subsection{Hybrid Tensor Network and Quantum Circuit Design}

A hybrid model with a Tensor Network and a quantum circuit is used to classify 28x28 MNIST images. The Tensor Network functions as a trainable feature extractor to compress the 784-dimensional data into 4 dimensions for classification by the quantum circuit.

There are several different types of tensor networks. For this study, the Matrix Product State (MPS) will be employed. The MPS, also referred to as a tensor train, is the simplest type of tensor network. In a MPS, tensors are contracted through virtual indices. The number of these indices is referred to as a bond dimension, denoted by $\chi$. A greater bond dimension indicates a greater amount of quantum entanglement that can be represented and therefore more representational power in the MPS. 
An N-dimensional input is mapped into a product state using the mapping  shown in Equation \ref{eq:tn_mapping}. This mapping for the MPS input is known as a feature map.

\begin{small}
\begin{equation}
x \rightarrow |\Psi\rangle = \begin{bmatrix}
           cos\left( \frac{\pi}{2}x_{1}\right) \\
           sin\left( \frac{\pi}{2}x_{1}\right) \\
         \end{bmatrix} \otimes \begin{bmatrix}
           cos\left( \frac{\pi}{2}x_{2}\right) \\
           sin\left( \frac{\pi}{2}x_{2}\right) \\ 
         \end{bmatrix} \otimes \hdots \otimes 
         \begin{bmatrix}
           cos\left( \frac{\pi}{2}x_{N}\right) \\
           sin\left( \frac{\pi}{2}x_{N}\right) \\ 
         \end{bmatrix}
\label{eq:tn_mapping}
\end{equation}
\end{small}

The MPS takes an input of size 784 (28 times 28) and outputs a n-length tensor. The output dimension of the MPS is a hyperparameter of the system that can be adjusted based on the problem at hand. This tensor output from the MPS is then encoded into a quantum circuit. $n$ dimensions are encoded onto $\frac{n}{2}$ qubits using  an $R_Y$ and $R_Z$ rotation on each qubit to encode two dimensions per qubit. Because the output of the MPS is not bounded, the arctangent of the input values are encoded for the rotations to keep inputs to the quantum circuit in the range of [-$\frac{\pi}{2}$,$\frac{\pi}{2}$]. After encoding, the circuit is run to get a quantum state fidelity measurement. This measurement is then mapped from [0.5,1] to [0,1] by subtracting 0.5 and multiplying by 2. The swap test may sometimes measure below 0.5 due to statistical error, so a ReLu layer is applied after the quantum circuit to prevent negative outputs. For multi-class classification, the ReLu layer is not used due to the presence of the softmax layer. If the output is below 0.5, the image is classified as 0, otherwise the image is classified as label 1. The quantum circuit has up to three types of layers: single-qubit unitary, dual-qubit unitary, and controlled-qubit unitary.

\begin{figure}[ht]			
	\centering
	\includegraphics[width=1\linewidth]{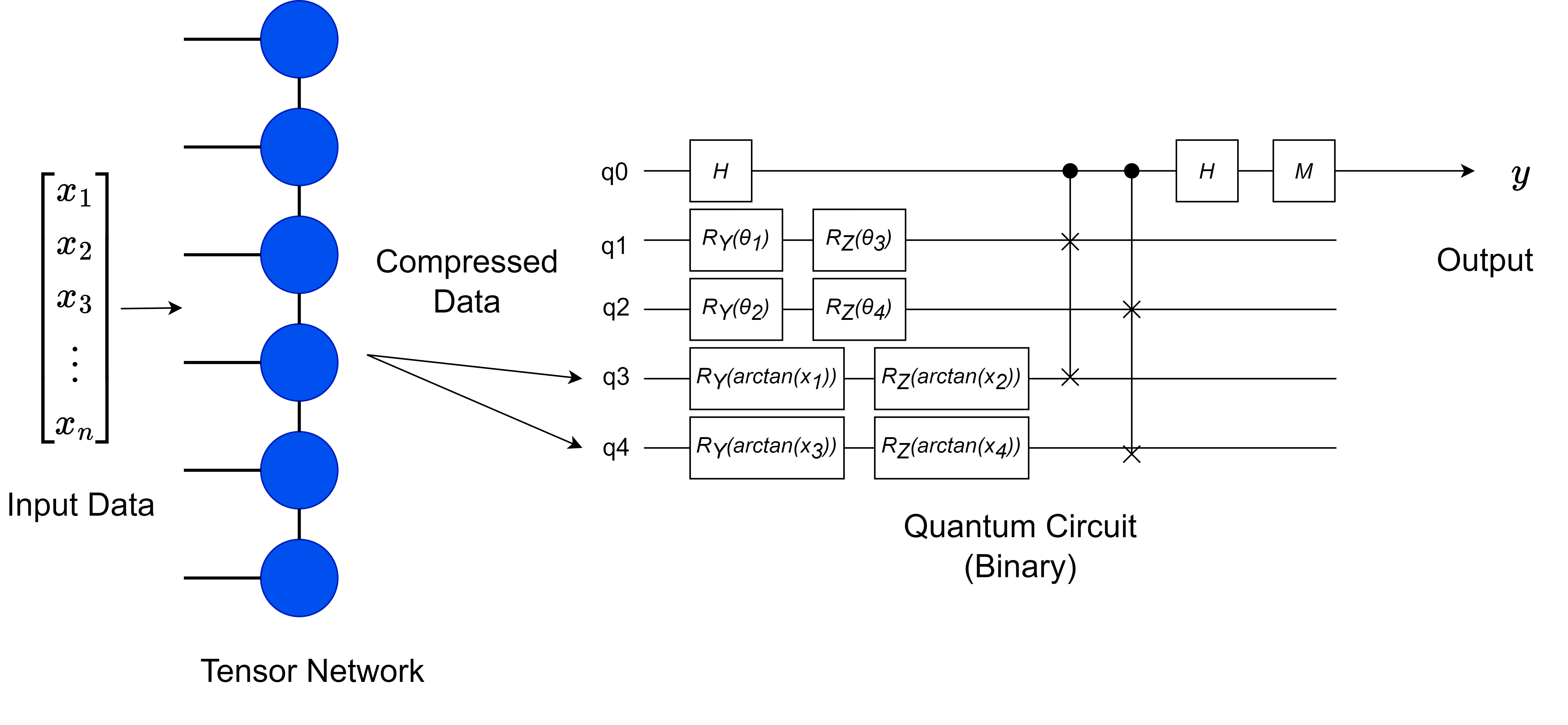}
    \caption{\sol~ Diagram (Binary)}
    \label{fig:sys_diagram_binary}
\end{figure}

\begin{figure}[ht]			
	\centering
	\includegraphics[width=1\linewidth]{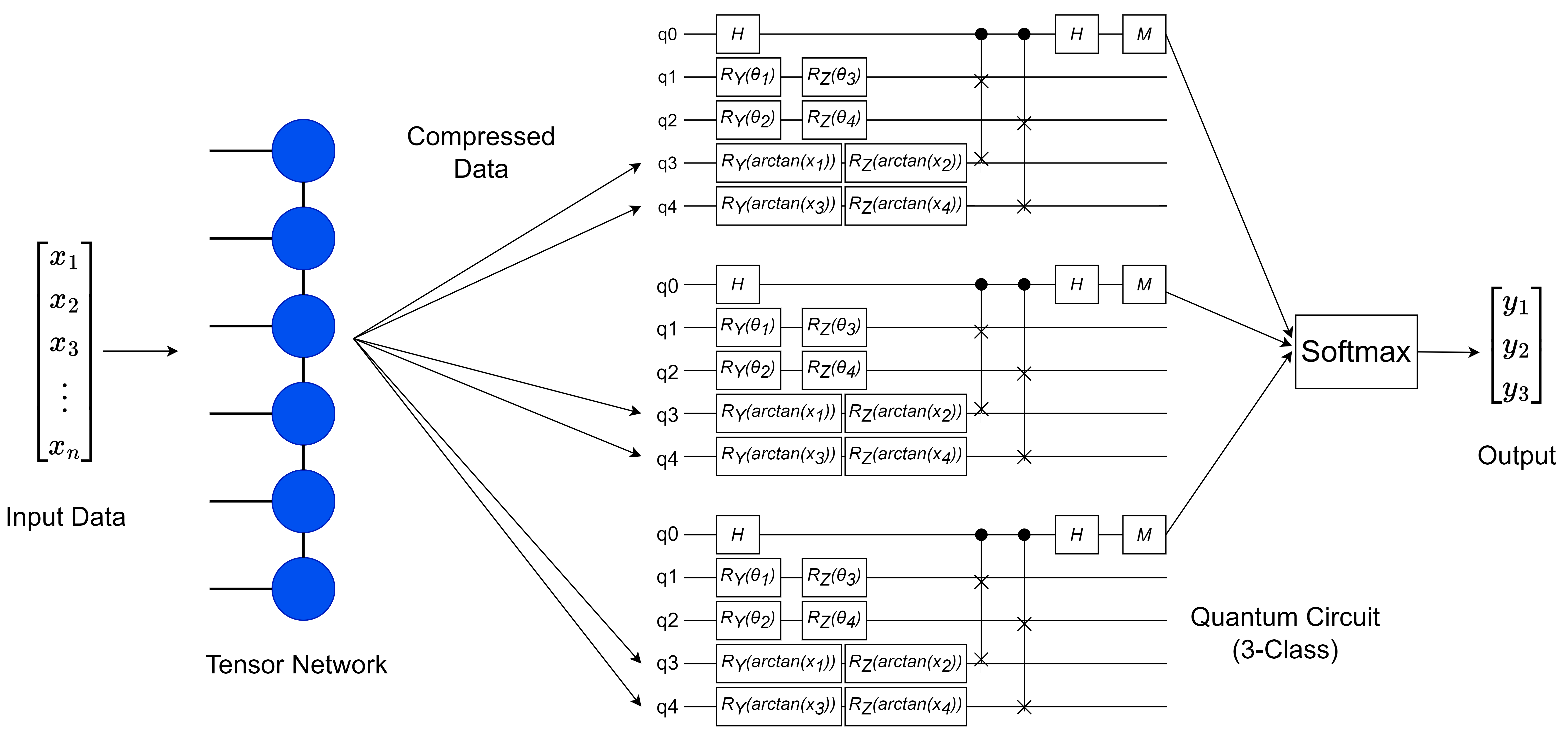}
    \caption{\sol~ Diagram (3-class)}
    \label{fig:sys_diagram_multiclass}
\end{figure}

For binary classification, a single quantum circuit is run. For n-class classification where $n > 2$, n quantum circuits with the same circuit design, but different parameters are run in parallel. The outputs of these circuits are then softmaxed to get probabilities for each class. The image is classified as the class with the highest probability. System diagrams for the binary and multi-class versions of this system can be seen in Figures \ref{fig:sys_diagram_binary} and \ref{fig:sys_diagram_multiclass}, respectively

This system can be trained all together at once rather than requiring a feature extractor to be pre-trained. 
The entire training algorithm is summarized in Algorithm  \ref{alg:cap}. First, the data is loaded as shown in Equation \ref{eq:tn_mapping} (Line 1). Lines 2-3 involve introducing training parameters set by the user at run time. The learning rate $\alpha$ indicates how large the updates to the system parameters should be during training. The network weights are initialized randomly. The number of epochs $\epsilon$ indicates how many times the network will be trained on the data set, $X$.  Line 6 represents the input data, $x$, being encoded into the tensor network. Line 7 represents the output of the tensor network being obtained through tensor contractions. Lines 8-23 represent the process by which each of the quantum parameters $\theta$ is updated. The output of the tensor network and the trainable quantum circuit parameters $\theta_{d}$ are all loaded into the quantum circuit with one of the parameters ($\theta$) either increased by $\frac{\pi}{2}$ ($\Delta_{fwd}$) and the SWAP test is performed. Then the parameters are reset, $\theta$ is decreased by $\frac{\pi}{2}$ ($\Delta_{bck}$), and the SWAP test is performed again.  The overall cost function of the network, $f(\theta_{d})$, is then obtained for the two adjusted parameter values and used to update $\theta$ as seen in  Line 22. After all of the quantum parameters have been updated, the parameters of the Tensor Network layer are updated as seen in Line 24. The quantum neural network is induced across all trained classes and the quantum state fidelity outputs are softmaxed. The class with the highest probability is returned as the classification. 

Algorithm  \ref{alg:cap} presents a hybrid training process that involves both classical and quantum ends, e.g., data loading and tensor networks on the classical side; quantum layers and measurements on the quantum side. The time and space complexity analysis should consider both quantum and classical resources. Due to the page limit and scope, we omit the theoretical algorithm analysis in this paper. 


\begin{algorithm}[ht]
\caption{\sol~ Algorithm}
\begin{algorithmic}[1]
\STATE Data set Loading Dataset: $(X \vert Class: Mixed)$
\STATE Distribute Dataset X By Class
\STATE Parameter Initialization:

\item [] Learning Rate : $\alpha = 10^{-4}$
\item [] Network Weights : $\theta_{d} = [$Rand Num between $0 - 1 \times \pi]$
\item [] epochs : $\epsilon = 40$
\item [] Dataset: $(X | Class = \omega)$ 
\item [] Qubit Channels: $Q = 2n_{X_{dim}}$
\item []

\FOR{$\zeta \in \epsilon$}
    \FOR{$x_{k} \in X$}
        \STATE Encode in Tensor Network $x \rightarrow \begin{bmatrix}
           cos\left( \frac{\pi}{2}x_{1}\right) \\
           sin\left( \frac{\pi}{2}x_{1}\right) \\
         \end{bmatrix} \otimes \begin{bmatrix}
           cos\left( \frac{\pi}{2}x_{2}\right) \\
           sin\left( \frac{\pi}{2}x_{2}\right) \\ 
         \end{bmatrix} \otimes \hdots \otimes
         \begin{bmatrix}
           cos\left( \frac{\pi}{2}x_{N}\right) \\
           sin\left( \frac{\pi}{2}x_{N}\right) \\ 
           \end{bmatrix}$  
\item []
        \STATE Perform Tensor contractions to get TN output 

\FOR{$\theta \in \theta_{d}$}   

    \STATE Perform Hadamard Gate on $Q_{0}$  
    \STATE Load $x_{k} \xrightarrow[\mathrm{Data Encoding}]{\mathrm{Quantum}} Q_{Q_{1}} \rightarrow Q_{count} $  

    \STATE Load $\theta_{d} \xrightarrow[\mathrm{Data Encoding}]{\mathrm{Quantum}} Q_{\frac{Q_{count}}{2}+1} + 1  \rightarrow \frac{Q_{count}}{2}+1 $ 
    \STATE Add $\frac{\pi}{2} \rightarrow \theta$ 
    \STATE $\Delta_{fwd} = (E_{Q_{0}}f(\theta_{d}))$
    \STATE CSWAP(Control Qubit = $Q_0$, Learned State Qubit, Data Qubit) 
    \STATE Measure $Q_0$ 
    \STATE Reset $Q_0$ to $|0 \rangle$
    \STATE Perform Hadamard Gate on $Q_0$ 
    \STATE Subtract $\frac{\pi}{2} \rightarrow \theta$ 
    \STATE CSWAP(Control Qubit = $Q_0$, Learned State Qubit, Data Qubit) 
    \STATE Measure $Q_0$  
    \STATE $\Delta_{bck} = (E_{Q_{0}}f(\theta_{d}))$
    \STATE $\theta = \theta - (0.5*(\Delta_{fwd}- \Delta_{bck})) \times \alpha$ 
        \ENDFOR
    \STATE Update Tensor Network parameters 
    \ENDFOR
\ENDFOR
\end{algorithmic}
\label{alg:cap}
\end{algorithm}

%% file: sources/evaluation.tex
\section{Evaluation}


We utilized Python 3.9 and the IBM Qiskit Quantum Computing simulator package to implement \sol~. The circuits were trained on NSF Cloudlab M510 nodes at the University of Utah datacenter. 
In our experiments, \sol~ is compared with state-of-the-art solutions listed below. 

\begin{itemize}
\item PCA-QuClassi~\cite{stein2022quclassi}: It is the predecessor of \sol. Instead of a collaborative quantum-classical training framework, it utilizes principal component analysis (PCA) to reduce the dimensions of the dataset. In our evaluations, we use PCA-5, PCA-7 and PCA-17 to denote its 5-qubit, 7-qubit and 17-qubit settings. Additionally, PCA-QuClassi has been compared with its different versions, including the Single Qubit Unitary Layer, Dual Qubit Unitary Layer and Entanglement Layer.

\item QuntumFlow~\cite{jiang2021co} (QF-pNet): It employs a co-design framework of quantum neural networks and utilizes downsampling to reduce the dimensions along with the amplitude encoding method.

\item TensorFlow Quantum~\cite{broughton2020tensorflow} (TFQ): The example codes provided by Tensorflow Quantum library are based on Cirq circuits and standard layer designs. 


\item DNN-Fair~\cite{fair}: A classical deep neural network for MNIST data may contain 1.2M parameters. For a more fair comparison, we construct a deep neural network with 3145 parameters. 

\end{itemize}


Furthermore, when comparing our \sol~ architecture to above-mentioned solutions in the literature of quantum deep learning, the MNIST dataset is a commonly used benchmark. 
MNIST comprises hand-written digits of resolution 28$\times$28, resulting in 784 dimensions. However, the evaluation data-encoding technique makes it impractical to perform experiments on near-term quantum computers and simulators due to the lack of qubits and computational complexity. As a result, we need to reduce the dimensionality to perform practical experiments. Therefore, it is necessary to reduce the dimensionality of the dataset. In our research, we have reduced the number of dimensions to 4 for binary experiments/simulations and 6 for multi-class evaluations. 
Besides original MNIST dataset, our evaluation involves two derived datasets, Fashion MNIST and Extended MNIST. We conducted both binary and multi-class experiments and evaluated them with simulators as well as IBM-Q quantum machines.

\subsection{Quantum Binary Classification}
\begin{figure}
\begin{subfigure}{.23\textwidth}
  \centering
  \includegraphics[width=0.80\linewidth]{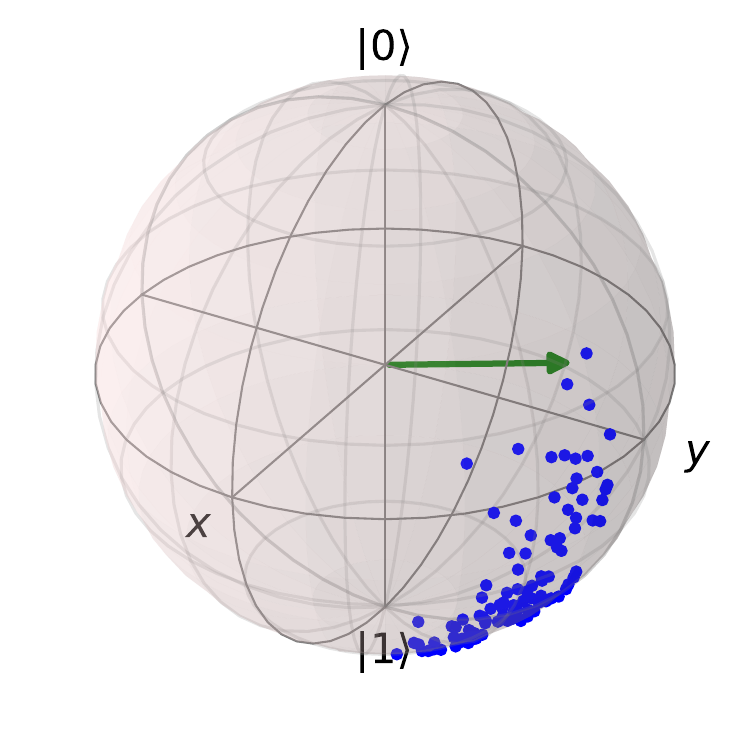}  
  \caption{Qubit 1 - 0 Epochs}
  \label{fig:sub-first}
\end{subfigure}
\begin{subfigure}{.23\textwidth}
  \centering
  \includegraphics[width=0.80\linewidth]{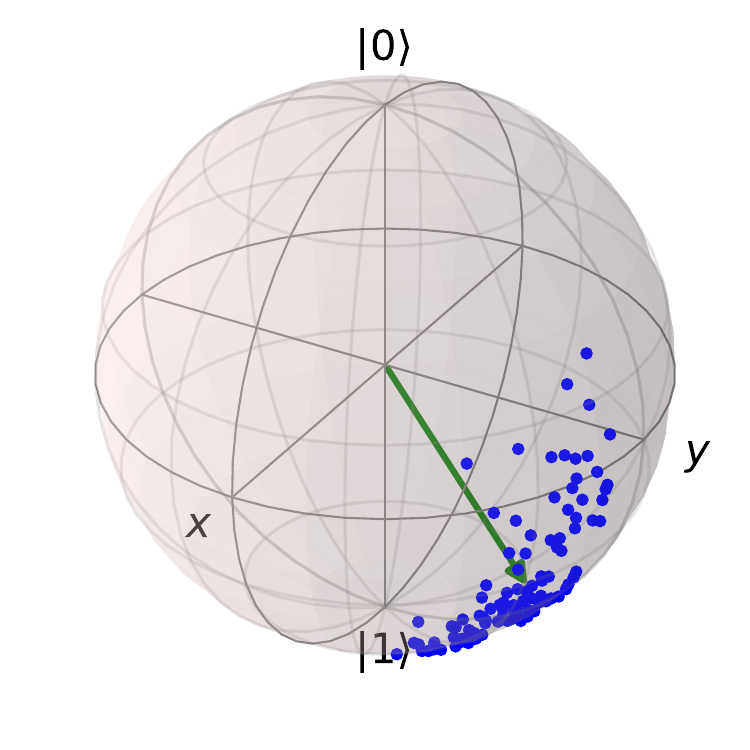}  
  \caption{Qubit 1 - 10 Epochs}
  \label{fig:sub-third}
\end{subfigure}
\caption{Identify 0 (Epoch 1 vs 10).}
\label{fig:quantum_state_evolution}
\end{figure}

In order to understand how our learning process works, we visualized the training process of identifying a 0 against a 6 by looking at the final state that is passed to the SWAP test. As illustrated in Fig.~\ref{fig:quantum_state_evolution}, an initial random quantum state is used to learn to classify 0 against 6. It is important to note that the state visualization does not account for potential learned entanglements, but serves as a visual aid to the learning process.
In Fig.~\ref{fig:quantum_state_evolution}, we can observe the evolution of the identifying state through epochs. The green arrows indicate the deep learning final state, and the blue points represent the training points. Initially, the identifying states are random, but they rotate and move towards the data, gradually minimizing the cost.

\begin{figure*}[ht]
\begin{subfigure}[b]{0.30\textwidth}			
	\centering
	\includegraphics[width=\textwidth]{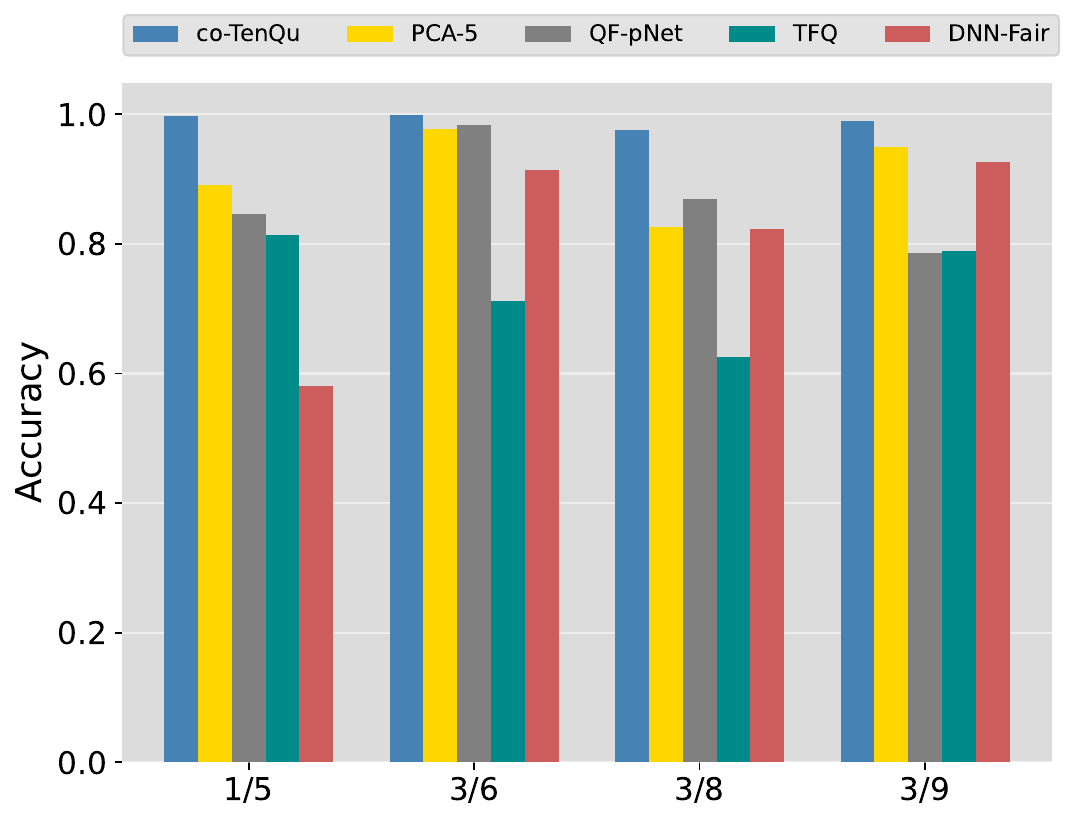}
    \caption{MNIST}
    \label{fig:mnist-binary}
\end{subfigure}
\begin{subfigure}[b]{0.33\textwidth}			
	\centering
	\includegraphics[width=\textwidth]{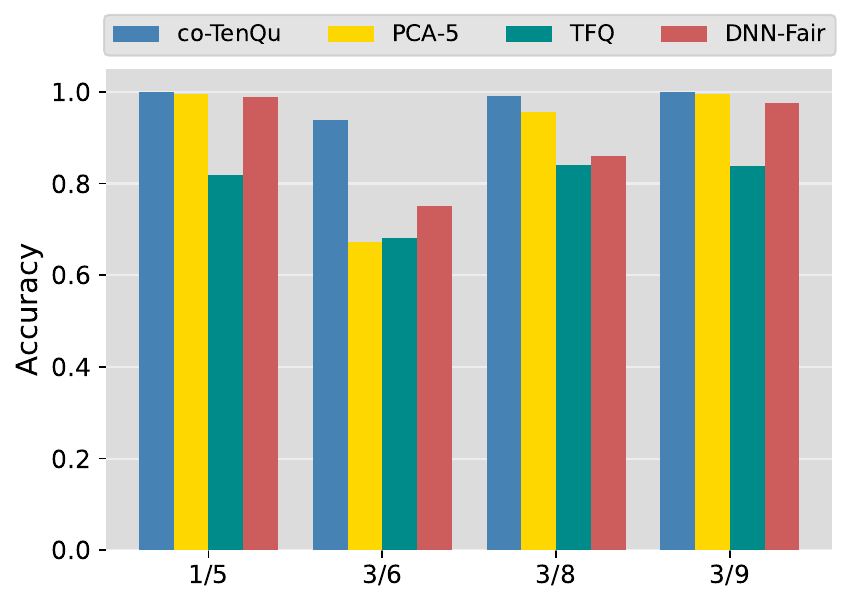}
    \caption{Fashion MNIST}
    \label{fig:fmnist-binary}
\end{subfigure}
\begin{subfigure}[b]{0.33\textwidth}			
	\includegraphics[width=\textwidth]{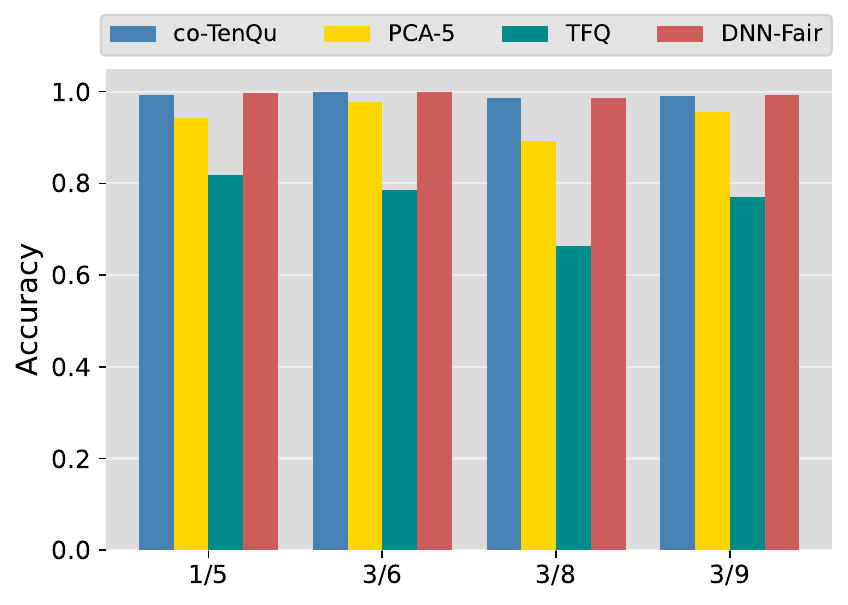}
	\captionsetup{justification=centering}
    \caption{Extended MNIST}
    \label{fig:emnist-binary}
\end{subfigure}
\caption{Binary Classifications with 5-qubit circuits for \sol~}
\label{fig:binary}
\end{figure*}

For binary classifications, we adopted popular digit combinations from the literature, specifically (1,5), (3,6), (3,8), and (3,9). The binary classification results are compared and visualized in Fig.\ref{fig:binary}. Clearly, \sol~ consistently outperforms all other solutions. For example, in the (1,5) classification with MNIST dataset (as shown on Fig.\ref{fig:mnist-binary}), it achieves the largest improvement of $41.72\%$ compared to classical deep neural networks, DNN-Fair (3145 parameters), with an accuracy of $99.79\%$. While classical DNN can achieve perfect accuracies on the MNIST dataset, it requires a much larger parameter size. By introducing 5 qubits, \sol~ is able to achieve better or similar performance with 49.54\% less parameters. 

When compared to quantum-based solutions with MNIST dataset, \sol~ outperforms others, with the largest margin achieved in the (3,8) and (3,9) classification, where we observe improvements of $35.07\%$ and $30.71\%$ over Tensorflow Quantum and QF-pNet. 
One noticeable thing is that, if we train Tensorflow Quantum with 17 qubits (verus 5 qubits), the accuracies increase substantially. For example, the accuracy boosted to from $71.25\%$ to $90.63\%$.
The primary difference between the designs is that \sol~ utilizes a quantum-state based evaluation function that can directly train the network on qubits and provide stable results.
\sol~ also outperformed its predecessor, PCA-QuClassi with MNIST dataset. While both employ a quantum-state based evaluation function, \sol~ incorporates a new trainable tensor network layer, allowing part of the training job to be completed on the classical part of the collaborative architecture.

A similar trend is discovered with both Fashion and Extended MNIST datasets as illustrated on Fig.\ref{fig:fmnist-binary} and Fig.\ref{fig:emnist-binary}. We can see that \sol~ outperforms all other solutions in compared 2-digit combinations. Comparing the results across 3 different datasets, TensorFlow Quantum's performance is not stable. For example, it achieves 62.58\%, 84.08\%, and 66.25\% for (3,8) classification that is a 21.50\% difference between datasets. With \sol~, however, the same value is 1.55\% with 97.65\%, 99.20\%, and 98.54\% for original, Fashion and Extended MNIST datasets respectively. 
\sol~ also beats PCA-QuClassi with 5-qubit setting (shown as PCA-5 on the figures) in all binary combinations with the largest gain, 26.58\%, observes at (3,6) Fashion MNIST (Fig.\ref{fig:fmnist-binary}). 
This is due to the fact that \sol~ utilizes the classical computational resource to partially complete training and pre-process the data for quantum parts.

\begin{figure}[ht]			
  \centering 
  \includegraphics[width=0.80\linewidth]{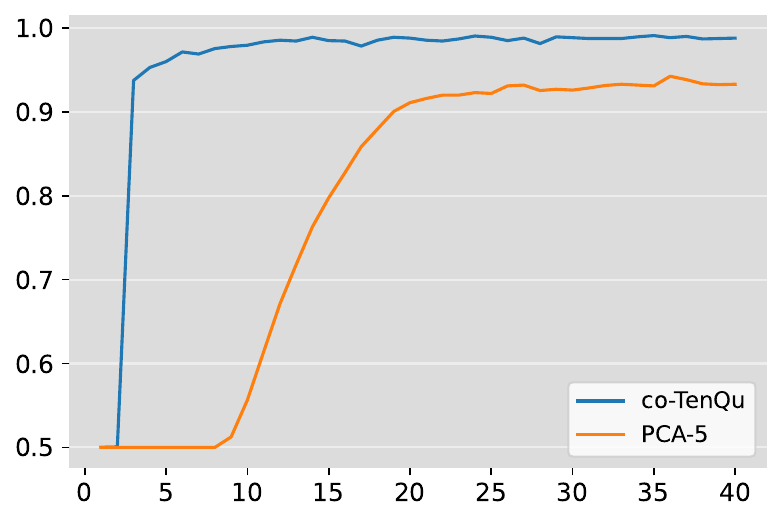}
    \caption{1/5 Extended MNIST Training}
    \label{fig:emnist-loss}
\vspace{-0.1in}    
\end{figure}

Furthermore, we find that \sol~ converges faster than PCA-QuClassi when taking a closer look at the training processes. Fig.\ref{fig:emnist-loss} presents the accuracy per each epoch of (1,5) classification on Extended MNIST dataset.  \sol~ reaches 93.75\% at its their epoch, after which it increases 5.10\% to 98.85\% at the $40^{th}$ epoch. Comparing with PCA-QuClassi with the 5 qubit setting, however, it records a 87.95\% accuracy at the $18^{th}$ epoch and climbs up to 93.30\% at the end, a 5.35\% increase. Given the training process, \sol~ converges significantly faster than PCA-QuClassi as it leverages trainable layers on the classical part.  

\subsection{Quantum Multi-class classification}

\begin{figure*}[ht]
\begin{subfigure}[b]{0.28\textwidth}			
	\centering
	\includegraphics[width=\textwidth]{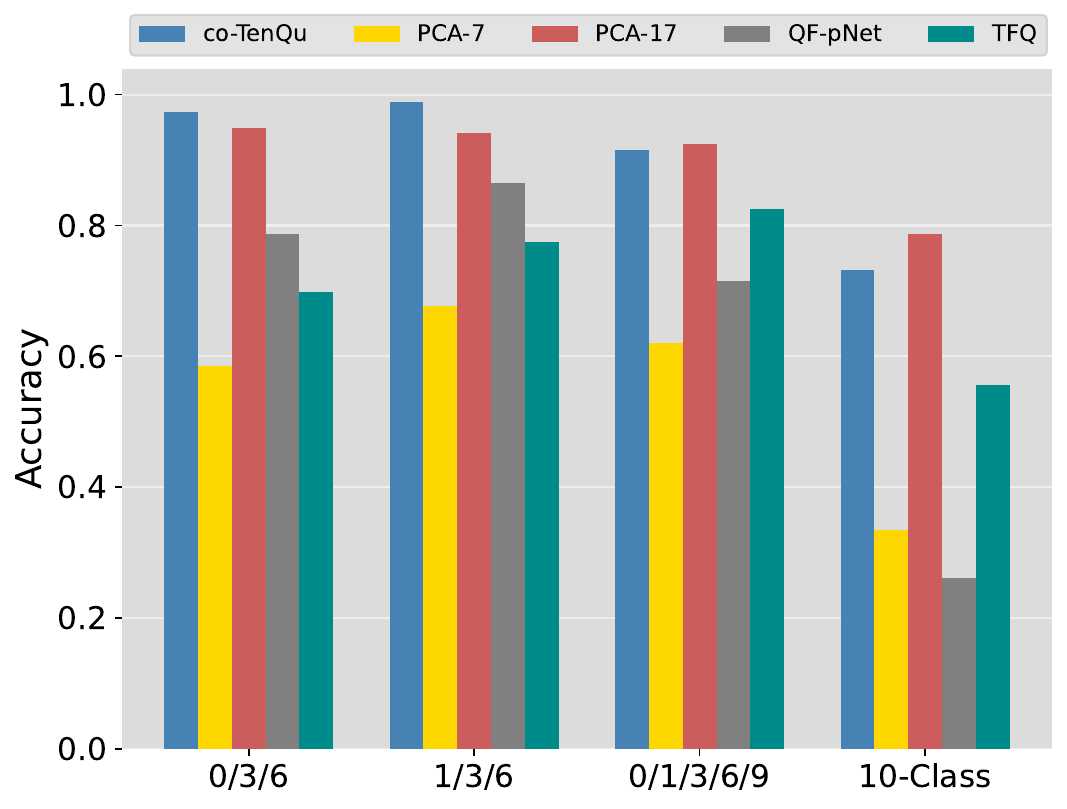}
    \caption{MNIST}
    \label{fig:mnist-multi}
\end{subfigure}
\begin{subfigure}[b]{0.33\textwidth}			
	\centering
	\includegraphics[width=\textwidth]{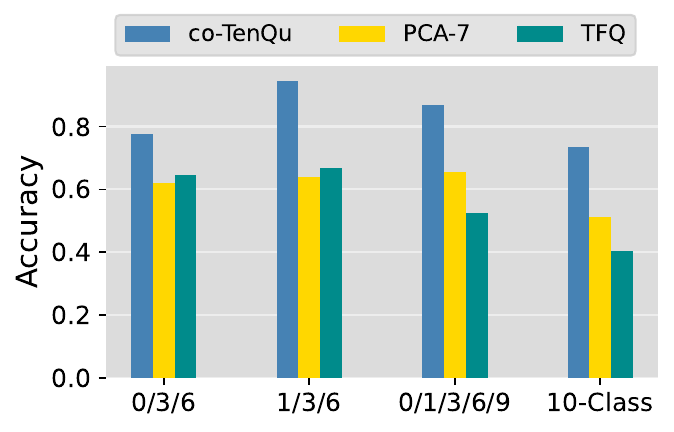}
    \caption{Fashion MNIST}
    \label{fig:fmnist-multi}
\end{subfigure}
\begin{subfigure}[b]{0.33\textwidth}			
	\includegraphics[width=\textwidth]{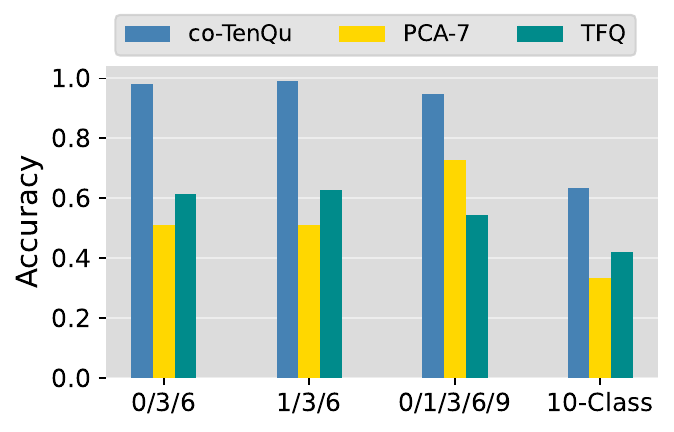}
	\captionsetup{justification=centering}
    \caption{Extended MNIST}
    \label{fig:emnist-multi}
\end{subfigure}
\caption{Multi-class Classifications with 7-qubit circuits for \sol}
\label{fig:multi}
\end{figure*}

Next, we evaluate our solution with multi-class classifications. In these experiments, \sol~ utilizes a 7-qubit setting.
The results demonstrate that \sol~ provides substantially better multi-class classification accuracies when comparing with the state-of-the-arts. 
With the multi-class classification, we select the popular digit combinations, (0,3,6), (1,3,6), (0,1,3,6,9) and 10-class, in the literature. The results are illustrated in Fig.\ref{fig:multi}. 
On the figure, we observe that \sol~ consistently outperforms other solutions. It achieves 97.39\%, 98.94\%, and 91.48\% for the first three multi-class experiments. PCA-QuClassi with the same 7-qubit setting records 58.55\%, 67.68\%, and 62.02\%. It demonstrates that \sol~ gains superior performance improvement, up to 66.3\%, by introducing the quantum-classical collaborative training architecture.
When increase the qubits utilization of PCA-QuClassi to the 17-qubit setting (shown as PCA-17 on the figures), its performance boosts to 94.91\%, 94.18\%, and 92.49\%  such that \sol~ wins the first two experiments, but fails the last one by 1\%. It further proves that \sol~ is able to achieve similar performance with 70.59\% less quantum resources (5 vs 17). Considering 10-class experiment, \sol~ performs significantly better PCA-QuClassi 7-qubit setting (73.21\% vs 33.41\%), but slightly worse than its 17-qubit version by 5\%. The reason lies in the fact that 17-qubit setting contains much more information for the training.

When comparing with QF-pNet, \sol~ improves the accuracies in all experiments. For example, \sol~ achieves 97.39\% and 98.94\% for (0,3,6) and (1,3,6), comparing with 78.70\% and 86.50\% obtained by QF-pNet, which leads to accuracy increases of 23.75\% and 14.38\%. In 5-class classification, \sol~ gains $19.92\%$ (91.48\% vs 71.56\%). As the number of classes increase, \sol~ outperforms QF-pNet by more than  $181.90\%$ (73.21\% vs 25.97\%) for 10-class classification. In QuantumFlow (QF-pNet), most of the training is done on the classical computer, where the traditional loss function is in use. With \sol, however, we employ a quantum-state based evaluation function that can fully utilize the qubits and a collaborative training architecture. 


We further compare \sol~ with PCA-QuClassi under the same 7-qubit setting with Fashion and Extended MNIST datasets. The same trend can be found on the Fig.\ref{fig:fmnist-multi} and Fig.\ref{fig:emnist-multi}, where \sol~ consistently outperforms its predecessor. It achieves the largest gain on (1,3,6) classification with Extended MNIST that is 99.06\% comparing with PCA-7's 50.90\%. \sol~ achieves stable performance on all 3-class and 5-class classifications across different datasets. For example, the accuracies for (0,3,6), (1,3,6) and (0,1,3,6,9) on Extended MNIST are 98.16\%, 99.06\%, and 94.88\%. With the 10-class job, the values drop to 73.40\% and 63.38\% for Fashion and Extended MNIST, respectively. However, \sol~ utilizes merely 7 qubits and performs much better, up to 1.90x, than PCA-QuClassi.

\subsection{Experiments on IBM-Q platform}
\begin{figure*}[!htb]
\centering
\begin{minipage}{0.7\textwidth}
\centering
	\includegraphics[width=1\linewidth]{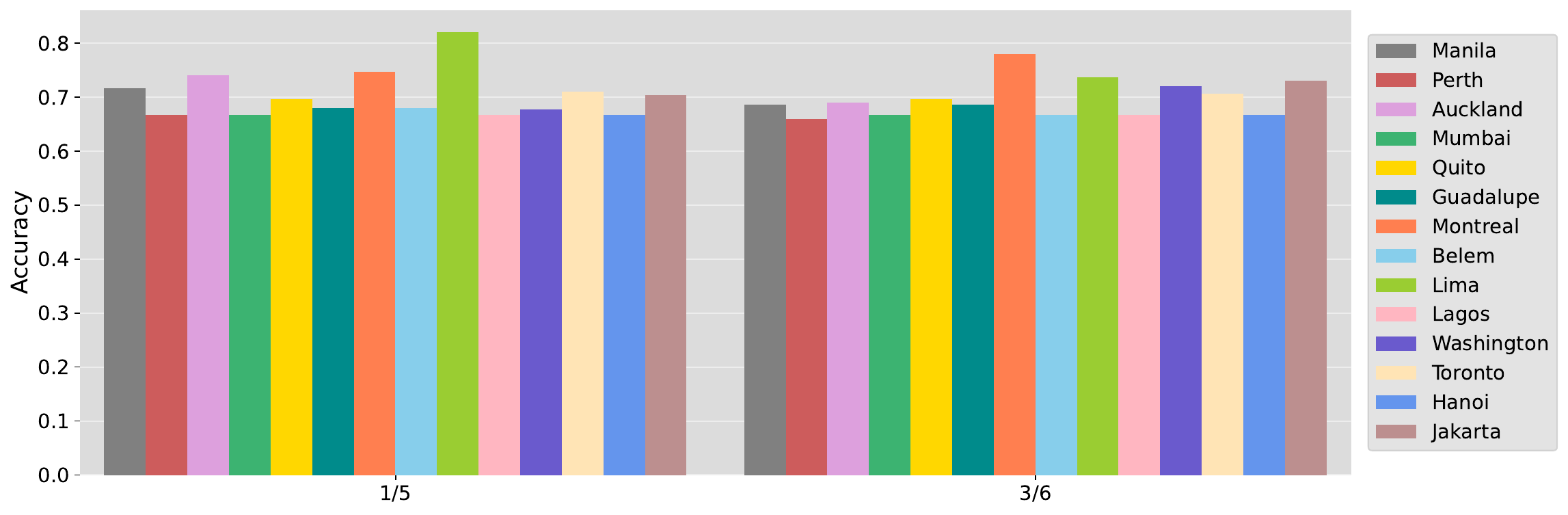}
    \caption{(1,5) and (3,6) MNIST Binary Classifications on IBM-Q Quantum}
    \label{fig:ibmq}
\end{minipage}
\begin{minipage}{0.29\textwidth}
\centering
\centering
	\includegraphics[width=0.79\linewidth]{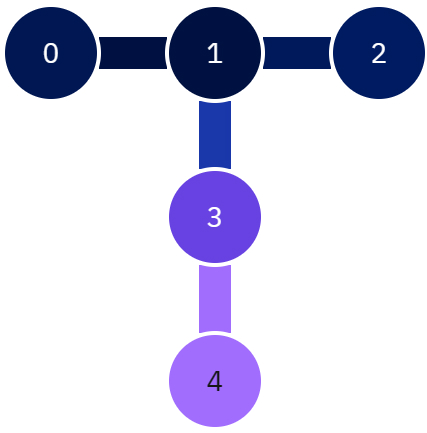}
    \caption{IBM-Q Lima Topology}
    \label{fig:lima}
\end{minipage}
\end{figure*}

As a proof of concept, we evaluate \sol~ on real quantum computers through the IBM-Q platform. 300 data points of the (1,5) and (3,6) MNIST experiments are submitted to 14 of IBM-Q's superconducting quantum computers.
Circuits are generated based off of a trained \sol{} network, whereby 300 circuits are submitted per machine in one job at 8192 shots each. The results are demonstrated in Fig.\ref{fig:ibmq}. 8 of the 14 machines generate a 66.67\% accuracy, which is the accuracy of the experiment for assuming all 0's (i.e. ground state). Variational parameters from simulation can perform poorly on real machines, with problems such as temporal drift and machine specific bias causing induction issues \cite{stein2022eqc}. Within tested machines, IBMQ-Lima achieved the best results, at 82.10\%. Lima's topology is drawn in Fig.\ref{fig:lima}, and has a Quantum Volume of 8, one of the lowest of IBM machines. This highlights the complexity that is predicting machine performance of quantum routines, and the implications that temporal drift has on learned parameters. Therefore, given sufficient resource, performance can be improved by optimizing the trained network locally, and finalizing training on the processor to learn the machine specific biases. 


\section{Discussion and Conclusion}
In this work, we propose \sol, a collaborative quantum-classic architecture for quantum neural networks. On the classical side, it utilizes a tensor network with trainable layers to preprocess the dataset to extract features and reduce the dimensionality. On the quantum part, it employs the quantum-state fidelity based cost function to train the model.  
Comparing to classical deep neural networks, \sol~ achieves 41.72\% accuracy improvement with a 49.54\% reduction in the parameter count.
Additionally, it outperforms other quantum-based solutions, up to 1.9 times, in multi-class classification. Furthermore, it records similar performance with 70.59\% less quantum resources. 
\sol~ represents a notable advancement in the realm of quantum deep learning. However, there remains considerable room for progress. Due to the limitations of current quantum machines, the existing solutions can only be evaluated on small dataset, such as MNIST.
In addition, the 10-class classification of MNIST resulted in a 73.21\% accuracy, which is relatively modest in comparison to classical counterparts. Although classical methods employ a higher number of parameters, they achieve accuracies approaching 100\%, which highlights the potential benefits that quantum computing could offer. 

Our future research will concentrate on extending the quantum-state fidelity based cost function and collaborative quantum-classical architecture to  other applications, such as quantum transformers and quantum natural language processing. 
 Additionally, exploring the low-qubit representation and its resilience to dynamic noises in the field of quantum-based learning warrants further investigation.